\documentstyle[psfig]{mn}	

\newcommand{\etal}{et~al.}			
\newcommand{\fsh}{4-Shooter}			
\newcommand{\kmsmpc}{km s$^{-1}$ Mpc$^{-1}$}	
\newcommand{\logten}{\mathop{{\rm log}_{10}}}	
\def\n{-\hfil}   				
\def\sn#1{\rlap{\hskip1pt $^{#1}$}}		
\newcommand{\th}{$^{\rm th}$}			
\newcommand{\WHzsr}{W~Hz$^{-1}$~sr$^{-1}$}	

\newcommand{\aap}{A\&A}
\newcommand{\aapss}{A\&AS}
\newcommand{\aj}{AJ}
\newcommand{\apj}{ApJ}
\newcommand{\apjl}{ApJ}
\newcommand{\apjss}{ApJS}

\newcommand{\mnras}{MNRAS}
\newcommand{\nat}{Nat}
\newcommand{\pasp}{PASP}

\newif\ifAMStwofonts
\AMStwofontstrue

\ifoldfss
  \ifCUPmtlplainloaded \else
    \NewTextAlphabet{textbfit} {cmbxti10} {}
    \NewTextAlphabet{textbfss} {cmssbx10} {}
    \NewMathAlphabet{mathbfit} {cmbxti10} {} 
    \NewMathAlphabet{mathbfss} {cmssbx10} {} 
  \fi
  \ifAMStwofonts
    \ifCUPmtlplainloaded \else
      \NewSymbolFont{upmath} {eurm10}
      \NewSymbolFont{AMSa} {msam10}
      \NewMathSymbol{\upi}     {0}{upmath}{19}
      \NewMathSymbol{\umu}     {0}{upmath}{16}
      \NewMathSymbol{\upartial}{0}{upmath}{40}
      \NewMathSymbol{\leqslant}{3}{AMSa}{36}
      \NewMathSymbol{\geqslant}{3}{AMSa}{3E}

       \let\le=\leqslant
       \let\ge=\geqslant
    \fi
  \fi
\fi 

\ifnfssone
  \newmathalphabet{\mathit}
  \addtoversion{normal}{\mathit}{cmr}{m}{it}
  \addtoversion{bold}{\mathit}{cmr}{bx}{it}
  \newmathalphabet{\mathbfit} 
  \addtoversion{normal}{\mathbfit}{cmr}{bx}{it}
  \addtoversion{bold}{\mathbfit}{cmr}{bx}{it}
  \newmathalphabet{\mathbfss} 
  \addtoversion{normal}{\mathbfss}{cmss}{bx}{n}
  \addtoversion{bold}{\mathbfss}{cmss}{bx}{n}
  \ifAMStwofonts
    \ifCUPmtlplainloaded \else
      \UseAMStwoboldmath
      \makeatletter
      \new@mathgroup\upmath@group
      \define@mathgroup\mv@normal\upmath@group{eur}{m}{n}
      \define@mathgroup\mv@bold\upmath@group{eur}{b}{n}
      \edef\UPM{\hexnumber\upmath@group}
      \new@mathgroup\amsa@group
      \define@mathgroup\mv@normal\amsa@group{msa}{m}{n}
      \define@mathgroup\mv@bold\amsa@group{msa}{m}{n}
      \edef\AMSa{\hexnumber\amsa@group}
      \makeatother
      \mathchardef\upi="0\UPM19
      \mathchardef\umu="0\UPM16
      \mathchardef\upartial="0\UPM40
      \mathchardef\leqslant="3\AMSa36
      \mathchardef\geqslant="3\AMSa3E

       \let\le=\leqslant
       \let\ge=\geqslant
    \fi
  \fi
\fi 

\ifnfsstwo
  \DeclareMathAlphabet{\mathbfit}{OT1}{cmr}{bx}{it}
  \SetMathAlphabet\mathbfit{bold}{OT1}{cmr}{bx}{it}
  \DeclareMathAlphabet{\mathbfss}{OT1}{cmss}{bx}{n}
  \SetMathAlphabet\mathbfss{bold}{OT1}{cmss}{bx}{n}
  \ifAMStwofonts
    \ifCUPmtlplainloaded \else
      \DeclareSymbolFont{UPM}{U}{eur}{m}{n}
      \SetSymbolFont{UPM}{bold}{U}{eur}{b}{n}
      \DeclareSymbolFont{AMSa}{U}{msa}{m}{n}
      \DeclareMathSymbol{\upi}{0}{UPM}{"19}
      \DeclareMathSymbol{\umu}{0}{UPM}{"16}
      \DeclareMathSymbol{\upartial}{0}{UPM}{"40}
      \DeclareMathSymbol{\leqslant}{3}{AMSa}{"36}
      \DeclareMathSymbol{\geqslant}{3}{AMSa}{"3E}

       \let\le=\leqslant
       \let\ge=\geqslant
    \fi
  \fi
\fi 

\ifCUPmtlplainloaded \else
  \ifAMStwofonts \else 
    \def\upi{\pi}
    \def\umu{\mu}
    \def\upartial{\partial}
  \fi
\fi

\title[LBDS Hercules: I. Multi-colour photometry]{The LBDS Hercules
sample of millijansky radio sources at 1.4~GHz: I. Multi-colour
photometry}

\author[Waddington et al.]{I. Waddington$^{1,2}$, R. A. Windhorst$^1$,
J. S. Dunlop$^2$, D. C.  Koo$^3$ and J. A. Peacock$^2$\\ $^1$
Department of Physics \& Astronomy, Arizona State University, PO Box
871504, Tempe, AZ 85287--1504, USA\\ $^2$ Institute for Astronomy,
University of Edinburgh, Royal Observatory, Blackford Hill, Edinburgh
EH9 3HJ, UK\\ $^3$ UCO/Lick Observatory, Department of Astronomy \&
Astrophysics, University of California, Santa Cruz, CA 95064, USA}

\date{2000 June 12 : Accepted for publication in MNRAS.}

\pagerange{\pageref{firstpage}--\pageref{lastpage}}
\pubyear{2000}

\begin{document}

\maketitle

\label{firstpage}

\begin{abstract}
The results are presented of an extensive programme of optical and
infrared imaging of radio sources in a complete subsample of the
Leiden--Berkeley Deep Survey.  The LBDS Hercules sample consists of 72
sources observed at 1.4~GHz, with flux densities $S_{1.4} \ge
1.0$~mJy, in a 1.2 deg$^2$ region of Hercules.  This sample is almost
completely identified in the $g$, $r$, $i$ and $K$ bands, with some
additional data available at $J$ and $H$.  The magnitude distributions
peak at $r\simeq 22$~mag, $K\simeq 16$~mag and extend down to $r\simeq
26$~mag, $K\simeq 21$~mag.  The $K$-band magnitude distributions for
the radio galaxies and quasars are compared with those of other radio
surveys.  At $S_{\rm 1.4~GHz}\la 1$~Jy, the $K$-band distribution does
not change significantly with radio flux density. The sources span a
broad range of colours, with several being extremely red ($r-K\ga 6$).
Although small, this is the most optically complete sample of
millijansky radio sources available at 1.4~GHz, and is ideally suited
to study the evolution of the radio luminosity function out to high
redshifts.
\end{abstract}



\section{Introduction}

Essentially all low-redshift, powerful ($\rm P_{\rm 1.4~GHz}\ge \rm
P^*_{\rm 1.4~GHz} = 10^{25}$ \WHzsr) radio galaxies are housed in
normal giant elliptical galaxies.  Thus radio selection has long been
regarded as providing an efficient method of identifying high-redshift
ellipticals.  However, very radio-powerful high-redshift sources have
proved virtually useless as probes of galaxy evolution due to the
contaminating effects of the active galactic nucleus (AGN) at
ultraviolet--infrared wavelengths (McCarthy \etal\ 1987; Tadhunter
\etal\ 1992; Best, Longair \& R\"ottgering 1996, 1997).  Given the
success of the Lyman-break technique for selecting high-redshift
galaxies on the basis of their optical colours (Steidel \etal\ 1996,
1999), it might be presumed that radio galaxies are no longer of any
importance for probing galaxy evolution.  In fact, nothing could be
further from the truth, since the identification of high-redshift
galaxies at optical wavelengths is by necessity biased towards the
bluest, most ultraviolet-active systems -- such methods cannot detect
old or dust-reddened sources.  What is required is an {\it optically
unbiased\/} selection method, such as radio selection at low flux
densities.  Unlike their high-power counterparts, faint radio sources
have only weak emission lines (Kron, Koo \& Windhorst 1985) and show
little evidence of an alignment effect (Dunlop \& Peacock 1993).

There are a number of radio surveys with millijansky flux density
limits at 1.4~GHz ($S_{1.4} \ga 1$~mJy).  This limit is sufficiently
{\it faint\/} that a large fraction of the sources potentially lie at
high redshifts, and yet sufficiently {\it bright\/} that objects
selected at this flux density are still predominantly radio galaxies
rather than low-redshift starbursts (Hopkins \etal\
2000)\nocite{Hopkins00}.  Some of the more recent surveys are the
FIRST survey (Becker, White \& Helfand 1995; White \etal\ 1997), the
NVSS (Condon \etal\ 1998), the VLA survey of the ELAIS regions
(Ciliegi \etal\ 1999), and the ATCA survey of the Marano Field
(Gruppioni \etal\ 1997).  In order to study the evolution of the hosts
of these radio sources, it is essential to identify optical/infrared
counterparts to a majority of the radio sources.  This has not yet
been achieved for any of these recent surveys.  This paper presents
essentially complete optical identifications for a subsample of the
Leiden--Berkeley Deep Survey (LBDS; Windhorst, van Heerde \& Katgert
1984a), together with infrared $K$-band observations of 85\% of the
optically identified sources.  A companion paper (Waddington \etal\
2000a; hereafter Paper II) will present the results of the
spectroscopic observations and photometric redshift estimates of these
sources.

The layout of the paper is as follows.  In section~2 the radio
observations and previous work using the LBDS are reviewed, and the
resolution corrections for the sources are recalculated.  In section~3
the optical identifications and photometry of sources in the Hercules
sample are presented.  The infrared observations are presented in
section~4.  Finally, section~5 is a discussion of the properties of
the sample as a whole and of a number of the individual sources, and
compares the $K$-band magnitude distribution to that of brighter radio
surveys.  Unless otherwise stated, a cosmology with $H_0 =
50$~\kmsmpc, $\Omega_0 = 1$ and $\Lambda = 0$ is assumed.

\section{The Leiden--Berkeley Deep Survey}

The LBDS began with a collection of multi-colour prime focus plates
that had been acquired with the 4-m Mayall Telescope at Kitt Peak for
the purpose of faint galaxy and quasar photometry \cite{Kron80,Koo82}.
Several high latitude fields were chosen in the selected areas SA28
(referred to as Lynx), SA57, SA68 and an area in Hercules.  The fields
were selected on purely optical criteria, such as high galactic
latitude ($b^{ II} \ga 35\degr$) and minimum H\thinspace{\sc i} column
density, resulting in a random area of sky from the radio perspective.
Nine of these fields were then surveyed with the 3-km Westerbork
Synthesis Radio Telescope (WSRT) at 21~cm (1.412~GHz) with a 12\farcs5
beam \cite{Windhorst84a}.  These observations yielded noise limited
maps with $\sigma=0.12$--0.28 mJy in 12~hours, with absolute
uncertainties in position of order 0\farcs4 and in flux of order
3--5\%.  A total of 471 sources were found, out of which a complete
sample of 306 sources was selected.  This complete sample is defined
by: (1) a peak signal-to-noise $S_{\rm p}/N \ge 5\sigma$,
corresponding to $S_{\rm p} \ge 0.6$--1.0 mJy; and (2) a limiting
primary beam attenuation of $-7$~dB, which corresponds to a radius of
0\fdg464, or an attenuation factor $A(r)\le 5$.  With this attenuation
radius, the WSRT field then covers the same area as the Mayall 4-m
plates.  Images were also made at the resolution of the 1.5-km WSRT
($\sigma=0.27$ mJy; Windhorst \etal\ 1984a)\nocite{Windhorst84a} and
at 50~cm (0.609~GHz) with the 3-km WSRT ($\sigma=0.48$ mJy; Windhorst
1984)\nocite{Windhorst84}.  The radio data for the two Hercules fields
are summarised in Table~\ref{radiotable}.


\begin{table*}
\begin{minipage}{13.0cm}
\caption{LBDS Hercules sample --  radio data.\label{radiotable}}
\begin{tabular}{lrrrccccccc}

Name\ \ \ \ \ \ \ \  & \multispan3{\hfil RA (J2000)\hfil } & $S_{1.4}$ (mJy) %
 & $S_{0.6}$ (mJy) & $\alpha^{1.4}_{0.6}$ & ${\cal R}$%
 & $\psi$ $\arcsec$ & $\phi$ $\degr$ & $W$ \\
& \multispan3{\hfil Dec (J2000)\hfil } & $\Delta S_{1.4}$ %
 & $\Delta S_{0.6}$ & $\Delta \alpha$ & & $\Delta\psi$
 & & \\
 & & & & & & & & & & \\

53W002  & 17 & 14 & 14.75 &  50.1 & 126.1 &  1.10 & R &  0.8 &  90 &  1.00 \\
        & 50 & 15 & 30.4  & (4.3) & (7.0) & (0.12) &  & (0.2) & &    \\
53W004  & 17 & 14 & 36.66 &  54.5 &  64.5 &  0.20 & U & \llap{$<$} 1.4 & 78 & 1.00 \\
        & 50 & 10 & 26.3  & (4.2) & (3.6) & (0.11) &  & - & &    \\
53W005  & 17 & 14 & 36.61 &   7.6 &  19.0 &  1.09 & E & 11.9 &  56 &  1.06 \\
        & 50 & 28 & 23.5  & (1.3) & (1.9) & (0.23) &  & - & &    \\
53W008  & 17 & 15 & 3.87  & 306.6 & 597.2 &  0.79 & R &  1.9 &  58 &  1.00 \\
        & 49 & 54 & 18.6  & (26.9) & (31.0) & (0.12) &  & (0.5) & &    \\
53W009\sn{\dagger} & 17 & 15 & 3.55  &  92.7 & 128.1 &  0.38 & R & 20.0 &  36 &  1.00 \\
        & 50 & 21 & 31.4  & (6.7) & (7.0) & (0.11) &  & (2.1) & &    \\
53W010\sn{\dagger}  & 17 & 15 & 5.48  &   8.1 &  15.1 &  0.73 & R & 11.0 & 100 &  1.07 \\
        & 50 & 10 & 19.1  & (0.8) & (1.2) & (0.16) &  & (4.5) & &    \\
53W011\sn{\dagger} & 17 & 15 & 6.29  &   3.5 & \llap{$<$}  2.8 & \llap{$<$} 0.28 & U & \llap{$<$} 8.8 & 0 &  1.64 \\
        & 49 & 55 & 46.6  & (0.7) &   -   &   -   &  & - & &    \\
53W012  & 17 & 15 & 9.14  &  47.6 &  67.3 &  0.41 & E & 20.0 &  49 &  1.00 \\
        & 50 &  0 & 26.0  & (3.8) & (3.6) & (0.11) &  & - & &    \\
53W013\sn{\dagger} & 17 & 15 & 14.47 &   3.7 & \llap{$<$} 2.7 & \llap{$<$$-$}0.39 & U & \llap{$<$} 10.5 & 125 & 1.70 \\
        & 49 & 53 & 55.6  & (0.8) &   -   &   -   &  & - & &    \\
53W014  & 17 & 15 & 18.00 &   5.3 & \llap{$<$} 2.7 & \llap{$<$$-$}0.81 & E &  2.9 & 169 &  1.21 \\
        & 50 &  0 & 31.2  & (0.7) &   -   &   -   &  & - & &    \\
53W015\sn{\dagger}  & 17 & 15 & 23.60 & 184.6 & 355.1 &  0.78 & R & 16.1 & 126 &  1.00 \\
        & 50 & 13 & 13.3  & (11.8) & (18.6) & (0.10) &  & (1.6) & &    \\
53W019  & 17 & 15 & 46.86 &   6.8 &  12.5 &  0.72 & E &  3.2 & 139 &  1.17 \\
        & 50 & 30 & 7.0   & (0.9) & (1.4) & (0.20) &  & - & &    \\
53W020  & 17 & 15 & 48.69 &   6.7 &  16.6 &  1.07 & R &  4.0 &  45 &  1.00 \\
        & 50 & 25 & 0.5   & (0.6) & (1.4) & (0.14) &  & (0.5) & &    \\
53W021  & 17 & 16 & 4.89  &   4.7 &  11.6 &  1.07 & E & 12.9 &  25 &  1.18 \\
        & 50 & 20 & 41.5  & (0.4) & (1.0) & (0.14) &  & - & &    \\
53W022\sn{*\dagger} & 17 & 16 & 9.26  &  11.8 &  16.9 &  0.43 & E & 21.8 &  67 &  1.26 \\
        & 50 & 24 & 18.0  & (1.0) & (1.4) & (0.14) &  & - & &    \\
53W023\sn{\dagger}  & 17 & 16 & 10.22 & 109.9 & 229.1 &  0.87 & R &  9.3 &  65 &  1.00 \\
        & 50 &  5 & 34.0  & (6.3) & (11.7) & (0.09) &  & (1.0) & &    \\
53W024  & 17 & 16 & 11.20 &  10.3 &  16.4 &  0.55 & U & \llap{$<$} 1.4 &  11 &  1.00 \\
        & 49 & 56 & 48.8  & (0.8) & (1.0) & (0.12) &  & - & &    \\
53W025\sn{\dagger}  & 17 & 16 & 22.71 &   1.1 & \llap{$<$} 2.7 & \llap{$<$} 1.03 & U & \llap{$<$} 8.7 &   0 & 10.68 \\
        & 50 & 16 & 57.8  & (0.2) &   -   &   -    &  & - & &    \\
53W026  & 17 & 16 & 28.38 &  21.1 &  39.2 &  0.74 & R &  3.5 & 138 &  1.00 \\
        & 49 & 47 & 12.8  & (2.0) & (2.1) & (0.13) &  & (4.0) & &    \\
53W027\sn{*\dagger} & 17 & 16 & 26.81 &   8.3 &  16.2 &  0.80 & E & 26.8 & 127 &  2.35 \\
        & 50 & 23 & 54.5  & (1.1) & (1.8) & (0.20) &  & - & &    \\
53W029  & 17 & 16 & 29.26 &  22.2 &  18.3 & \llap{$-$}0.23 & U & \llap{$<$} 1.4 &   0 &  1.00 \\
        & 50 & 20 & 22.1  & (1.3) & (1.2) & (0.10) &  & - & &    \\
53W030  & 17 & 16 & 43.01 &   1.4 & \llap{$<$} 2.3 & \llap{$<$} 0.60 & U & \llap{$<$} 1.0 &   0 &  3.63 \\
        & 50 &  9 & 40.3  & (0.2) &   -   &   -    &  & - & &    \\
53W031  & 17 & 16 & 46.32 & 116.5 & 210.2 &  0.70 & R &  4.1 & 111 &  1.00 \\
        & 49 & 56 & 44.4  & (6.9) & (10.6) & (0.09) &  & (0.5) & &    \\
53W032\sn{*\dagger} & 17 & 16 & 46.99 &  10.5 &  20.5 &  0.80 & E & 22.4 & 155 &  1.23 \\
        & 49 & 48 & 51.5  & (2.1) & (1.6) & (0.26) &  & - & &    \\
53W034\sn{*\dagger} & 17 & 16 & 54.16 &  10.9 &  25.3 &  1.00 & E & 40.2 &  38 &  1.88 \\
        & 50 &  0 & 39.2  & (1.4) & (1.8) & (0.17) &  & - & &    \\
53W035  & 17 & 16 & 55.76 &   4.4 &   3.0 & \llap{$-$}0.44 & U & \llap{$<$} 1.3 & 133 &  1.46 \\
        & 50 & 18 & 39.5  & (0.4) & (0.6) & (0.27) &  & - & &    \\
53W036\sn{\dagger}  & 17 & 16 & 56.46 &   3.2 &   9.0 &  1.24 & U & \llap{$<$} 6.5 &   0 &  1.29 \\
        & 50 & 29 & 2.7   & (0.3) & (0.9) & (0.18) &  & - & &    \\
53W037  & 17 & 16 & 59.72 &   6.6 &  16.3 &  1.07 & E &  3.6 &  66 &  1.00 \\
        & 50 & 19 & 16.2  & (0.4) & (1.2) & (0.12) &  & - & &    \\
53W039  & 17 & 17 & 2.00 &   3.4 &   6.8 &  0.82 & U & \llap{$<$} 11.4 & 2 &  1.57 \\
        & 50 & 25 & 29.0  & (0.4) & (1.0) & (0.22) &  & - & &    \\
53W041  & 17 & 17 & 19.19 &   9.4 &  19.8 &  0.88 & U & \llap{$<$} 1.4 & 151 &  1.01 \\
        & 49 & 49 & 18.3  & (0.9) & (1.1) & (0.14) &  & - & &    \\
53W042  & 17 & 17 & 24.12 &   6.6 &  16.1 &  1.07 & R &  1.3 &   0 &  1.26 \\
        & 49 & 48 & 26.1  & (0.8) & (1.0) & (0.17) &  & (0.5) & &    \\

\end{tabular}
\end{minipage}
\end{table*}

\begin{table*}
\begin{minipage}{13.0cm}
\contcaption{}
\begin{tabular}{lrrrccccccc}

Name\ \ \ \ \ \ \ \  & \multispan3{\hfil RA (J2000)\hfil } & $S_{1.4}$ (mJy) %
 & $S_{0.6}$ (mJy) & $\alpha^{1.4}_{0.6}$ & ${\cal R}$%
 & $\psi$ $\arcsec$ & $\phi$ $\degr$ & $W$ \\
& \multispan3{\hfil Dec (J2000)\hfil } & $\Delta S_{1.4}$ %
 & $\Delta S_{0.6}$ & $\Delta \alpha$ & & $\Delta\psi$
 & & \\
 & & & & & & & & & & \\

53W043  & 17 & 17 & 29.71 &   2.7 &   9.6 &  1.52 & E &  2.7 &   0 &  1.65 \\
        & 50 & 18 & 52.8  & (0.3) & (0.9) & (0.17) &  & - & &    \\
53W044  & 17 & 17 & 36.90 &   1.8 &   3.8 &  0.92 & R &  2.0 &   0 &  3.47 \\
        & 50 &  3 &  4.6  & (0.3) & (0.5) & (0.23) &  & (0.5) & &    \\
53W045  & 17 & 17 & 41.47 &   1.5 & \llap{$<$} 2.4 & \llap{$<$} 0.58 & U & \llap{$<$} 1.0 &  17 &  5.32 \\
        & 50 & 15 & 44.0  & (0.3) &   -   &   -    &  & - & &    \\
53W046  & 17 & 17 & 53.27 &  63.1 & 112.6 &  0.69 & R &  3.2 & 143 &  1.03 \\
        & 50 &  7 & 51.4  & (3.2) & (5.7) & (0.09) &  & (0.5) & &    \\
53W047  & 17 & 18 &  7.89 &  23.9 &  42.0 &  0.67 & R &  1.4 &  94 &  1.00 \\
        & 50 & 22 & 44.6  & (1.6) & (2.3) & (0.10) &  & (0.5) & &    \\
53W048  & 17 & 18 & 11.17 &  11.5 &  22.6 &  0.81 & R &  1.3 &  90 &  1.00 \\
        & 50 & 24 &  1.1  & (1.1) & (1.7) & (0.14) &  & (0.5) & &    \\
53W049  & 17 & 18 & 11.26 &  95.1 & 188.9 &  0.81 & E & 30.9 &  78 &  1.00 \\
        & 50 & 33 & 13.8  & (8.0) & (10.2) & (0.12) &  & - & &    \\
53W051\sn{\dagger}  & 17 & 18 & 30.19 & 141.6 & 294.3 &  0.87 & R & 19.6 &  30 &  1.00 \\
        & 49 & 48 & 32.5  & (8.3) & (14.8) & (0.09) &  & (1.6) & &    \\
53W052  & 17 & 18 & 34.18 &   8.6 &  16.1 &  0.74 & R &  2.2 &   9 &  1.03 \\
        & 49 & 58 & 53.0  & (0.6) & (1.0) & (0.10) &  & (0.5) & &    \\
53W054A\sn{\dagger} & 17 & 18 & 47.12 &   3.9 &   2.8 & \llap{$-$}0.39 & U & \llap{$<$} 12.5 &  50 &  1.57 \\
        & 49 & 45 & 49.6  & (0.6) & (0.4) & (0.24) &  & - & &    \\
53W054B\sn{\dagger} & 17 & 18 & 49.87 &   3.0 &   2.1 & \llap{$-$}0.42 & U & \llap{$<$} 12.5 &  50 &  1.73 \\
        & 49 & 46 & 12.4  & (0.4) & (0.3) & (0.25) &  & - & &    \\
53W057  & 17 & 19 &  7.33 &   2.9 & \llap{$<$} 2.2 & \llap{$<$$-$}0.36 & R &  1.0 & 122 &  3.33 \\
        & 49 & 45 & 45.1  & (0.6) &   -   &   -    &  & (0.5) & &    \\
53W058\sn{\dagger}  & 17 & 19 & 19.12 &   1.4 & \llap{$<$} 2.2 & \llap{$<$} 0.52 & U & \llap{$<$} 11.3 &  45 &  6.68 \\
        & 49 & 57 & 47.8  & (0.3) &   -   &    -   &  & - & &    \\
53W059  & 17 & 19 & 20.47 &  18.7 &  40.0 &  0.90 & U & \llap{$<$} 3.0 &  97 &  1.00 \\
        & 50 &  0 & 19.2  & (1.0) & (2.1) & (0.09) &  & - & &    \\
53W060  & 17 & 19 & 25.08 &   9.7 &  21.1 &  0.93 & U & \llap{$<$} 1.4 &  66 &  1.02 \\
        & 49 & 28 & 44.6  & (1.4) & (1.3) & (0.19) &  & - & &    \\
53W061\sn{\dagger}  & 17 & 19 & 27.37 &   2.6 & \llap{$<$} 2.3 & \llap{$<$$-$}0.15 & R & 12.3 &  72 &  2.35 \\
        & 49 & 43 & 59.7  & (0.4) &   -   &    -   &  & (5.4) & &    \\
53W062  & 17 & 19 & 31.97 &   1.7 & \llap{$<$} 2.2 & \llap{$<$} 0.28 & U & \llap{$<$} 1.4 &   0 &  6.46 \\
        & 49 & 59 &  6.5  & (0.3) &   -   &   -    &  & - & &    \\
53W065  & 17 & 19 & 40.11 &   5.3 &  14.5 &  1.21 & U & \llap{$<$} 1.4 &   0 &  1.00 \\
        & 49 & 57 & 39.3  & (0.4) & (0.9) & (0.11) &  & - & &    \\
53W066  & 17 & 19 & 42.97 &   4.1 &   8.8 &  0.91 & U & \llap{$<$} 1.4 &   0 &  1.12 \\
        & 50 &  1 &  4.5  & (0.3) & (0.7) & (0.13) &  & - & &    \\
53W067  & 17 & 19 & 51.63 &  23.2 &  45.8 &  0.81 & E & 12.6 & 101 &  1.00 \\
        & 50 & 10 & 55.4  & (1.4) & (2.5) & (0.10) &  & - & &    \\
53W068  & 17 & 19 & 59.72 &   3.9 &   5.1 &  0.33 & U & \llap{$<$} 1.0 &  31 &  1.26 \\
        & 49 & 36 &  9.5  & (0.5) & (0.6) & (0.20) &  & - & &    \\
53W069  & 17 & 20 &  2.49 &   3.7 &   7.8 &  0.87 & R &  1.6 &   0 &  1.18 \\
        & 49 & 44 & 51.1  & (0.3) & (0.9) & (0.17) &  & (0.5) & &    \\
53W070  & 17 & 20 &  6.18 &   2.6 & \llap{$<$} 2.5 & \llap{$<$$-$}0.04 & U & \llap{$<$} 1.4 & 107 &  1.96 \\
        & 50 &  6 &  1.7  & (0.3) &   -   &   -    &  & - & &    \\
53W071  & 17 & 20 & 11.34 &   2.8 &   9.3 &  1.43 & U & \llap{$<$} 1.4 & 132 &  1.89 \\
        & 50 & 17 & 17.1  & (0.6) & (0.9) & (0.26) &  & - & &    \\
53W072  & 17 & 20 & 29.56 &   6.6 &   7.6 &  0.17 & U & \llap{$<$} 1.4 &  25 &  1.13 \\
        & 50 & 22 & 37.6  & (1.3) & (1.0) & (0.29) &  & - & &    \\
53W075  & 17 & 20 & 42.41 &  96.1 & 185.3 &  0.78 & U & \llap{$<$} 1.2 &   0 &  1.00 \\
        & 49 & 43 & 49.0  & (6.0) & (9.6) & (0.10) &  & - & &    \\
53W076\sn{\dagger}  & 17 & 20 & 56.09 &   1.4 & \llap{$<$} 3.0 & \llap{$<$} 0.89 & R &  1.0 &   0 &  3.99 \\
        & 49 & 40 &  1.1  & (0.3) &   -   &   -    &  & (0.5) & &    \\
53W077\sn{\dagger}  & 17 & 21 &  0.21 &   7.8 &  16.1 &  0.87 & R & 16.8 &  82 &  1.54 \\
        & 49 & 48 & 31.4  & (0.8) & (1.6) & (0.17) &  & (4.2) & &    \\
53W078  & 17 & 21 & 18.20 &   2.0 & \llap{$<$} 3.1 & \llap{$<$} 0.53 & R &  0.7 &   0 &  3.38 \\
        & 50 &  3 & 35.2  & (0.4) &   -   &   -    &  & (0.5) & &    \\
53W079  & 17 & 21 & 22.79 &  13.3 &  13.9 &  0.05 & U & \llap{$<$} 1.4 &  39 &  1.00 \\
        & 50 & 10 & 31.3  & (1.0) & (1.3) & (0.14) &  & - & &    \\
53W080  & 17 & 21 & 37.53 &  27.6 &  54.2 &  0.80 & E & 10.4 &  35 &  1.00 \\
        & 49 & 55 & 37.2  & (1.8) & (3.0) & (0.10) &  & (0.5) & &    \\

\end{tabular}
\end{minipage}
\end{table*}

\begin{table*}
\begin{minipage}{13.0cm}
\contcaption{}
\begin{tabular}{lrrrccccccc}

Name\ \ \ \ \ \ \ \  & \multispan3{\hfil RA (J2000)\hfil } & $S_{1.4}$ (mJy) %
 & $S_{0.6}$ (mJy) & $\alpha^{1.4}_{0.6}$ & ${\cal R}$%
 & $\psi$ $\arcsec$ & $\phi$ $\degr$ & $W$ \\
& \multispan3{\hfil Dec (J2000)\hfil } & $\Delta S_{1.4}$ %
 & $\Delta S_{0.6}$ & $\Delta \alpha$ & & $\Delta\psi$
 & & \\
 & & & & & & & & & & \\

53W081  & 17 & 21 & 37.90 &  12.2 &  24.7 &  0.84 & U & \llap{$<$} 1.4 &   0 &  1.00 \\
        & 49 & 57 & 58.0  & (0.8) & (1.6) & (0.11) &  & - & &    \\
53W082  & 17 & 21 & 37.70 &   2.0 &   6.5 &  1.41 & U & \llap{$<$} 1.4 &   0 &  2.28 \\
        & 50 &  8 & 27.7  & (0.4) & (1.0) & (0.29) &  & - & &    \\
53W083  & 17 & 21 & 49.00 &   5.0 &   9.0 &  0.70 & U & \llap{$<$} 1.4 &   0 &  1.00 \\
        & 50 &  2 & 40.1  & (0.5) & (1.1) & (0.19) &  & - & &    \\
53W085  & 17 & 21 & 52.54 &   4.3 &  12.8 &  1.29 & U & \llap{$<$} 1.4 &   0 &  1.04 \\
        & 49 & 54 & 34.5  & (0.4) & (1.1) & (0.16) &  & - & &    \\
53W086  & 17 & 21 & 57.74 &   4.9 &   6.7 &  0.35 & R &  1.3 & 124 &  1.52 \\
        & 49 & 53 & 34.5  & (0.7) & (1.1) & (0.26) &  & (0.5) & &    \\
53W087  & 17 & 21 & 59.11 &   5.8 &  15.8 &  1.18 & E &  2.9 & 106 &  1.02 \\
        & 50 &  8 & 43.7  & (0.8) & (1.6) & (0.20) &  & - & &    \\
53W088  & 17 & 21 & 58.89 &  14.9 &  13.7 & \llap{$-$}0.10 & U & \llap{$<$} 1.4 &  22 &  1.00 \\
        & 50 & 11 & 53.5  & (1.4) & (1.2) & (0.15) &  & - & &    \\
53W089  & 17 & 22 &  1.17 &   2.5 &   7.3 &  1.29 & E &  3.2 & 178 &  2.12 \\
        & 50 &  6 & 53.2  & (0.5) & (1.1) & (0.30) &  & - & &    \\
53W090\sn{\dagger}  & 17 & 22 & 23.89 &   2.1 & \llap{$<$} 4.1 & \llap{$<$} 0.83 & U & \llap{$<$} 9.3 &   0 &  2.50 \\
        & 49 & 56 & 44.5  & (0.4) &   -   &   -    &  & - & &    \\
53W091  & 17 & 22 & 32.73 &  22.1 &  66.0 &  1.30 & E &  4.0 &   0 &  1.00 \\
        & 50 &  6 &  1.8  & (2.0) & (3.9) & (0.13) &  & - & &    \\

\end{tabular}

\medskip

$*$ The source has multiple radio components -- the coordinates
correspond to the average position of all the components.

$\dagger$ WSRT data only \cite{Windhorst84a} -- all other sources have
VLA positions and morphologies \cite{Oort87,Oort88a}.  All coordinates
have been converted to a J2000 equinox.

\end{minipage}
\end{table*}

Following this selection of the radio sample, the photographic plates
were searched for optical counterparts to the radio sources
(Windhorst, Kron \& Koo 1984b)\nocite{Windhorst84b}.  Large
astrographic plates were used to measure the positions of 30--40
secondary standards based on the positions of AGK-3 astrometric
standard stars.  These secondary standards were then used to calibrate
the Mayall prime focus plates for each field in the survey.  The radio
and optical position errors combine to give an error ellipse around
the co-ordinates of the radio source within which one expects to find
the optical identification of the source.  For the complete sample,
171 of the 306 radio sources were identified in this way, to $U^+$,
$J^+ \simeq 23.5$~mag and $F^+$, $N^+ \simeq 21.5$--22.5~mag.  The
expected number of real identifications was 53\% for the whole survey,
while for the Hercules fields the identification percentage was
somewhat higher at 65\% (47 out of 72 sources).  As will be discussed
in Paper~II, there is some evidence for large-scale structure in
Hercules, which may explain the higher identification rate in these
two fields.

Magnitudes for each of these objects were measured from digital scans
of the plates \cite{Kron85}.  Absolute calibrations were made for most
of the fields using photoelectric observations of stars in the fields
and transforming the derived $UBV$ zero-points to the photographic
$U^+J^+F^+N^+$ system.  For the two Hercules fields, however, no
photoelectric zero-points were available. They had to be estimated by
assuming that the sensitivity of the Hercules plates was the same as
for plates (with the same emulsion and filter) taken of another field
for which photometric calibrations were available.  The resulting
errors in the photometry (0.2--0.3 mag) are thus larger for Hercules
than the rest of the sample.  Redshifts were measured for
approximately 60 of the LBDS identifications, of which 16 were in
Hercules \cite{Kron85}.

One of the major limiting factors in using the LBDS for statistical
studies of radio source evolution has been the incomplete optical
identification of the sample and the limited number of redshifts
available.  To obtain complete identification and spectroscopic data
on the whole 306-source survey was a major long-term project, due to
the size of the survey area (several square degrees), so here
attention is focused on a subsample of 72 sources.  The two Hercules
fields were chosen for this detailed study because they had the
largest number of optical identifications already available.  In
addition, the highest redshift object yet known in the LBDS (53W002 at
$z=2.390$) was found in these fields.  Optical and infrared
observations have been acquired over many years and are presented here
for the first time in their entirety.  First, some of the relevant
highlights of previous work in the LBDS will be reviewed and the radio
observations of Hercules will be summarised.

\subsection{Previous work in the LBDS}

Results from the photographic identifications were presented by Kron
\etal\ (1985).\nocite{Kron85} About 20\% of the identifications were
quasars and there were two stars. The galaxies were separated into
bright ($F^+<18$) and faint ($18<F^+<21.5$), red ($J^+-F^+>1.2$--2.8)
and blue ($J^+-F^+<1.2$--2.8) classes.  The red radio galaxies (both
bright \& faint classes) were identified as giant ellipticals with
extended radio emission -- ``classical'' powerful radio galaxies.
The bright, blue galaxies were exclusively spirals with unresolved
radio structure and a very narrow range in power $\logten P_{1.4} =
22.1 \pm 0.1$.  The faint, blue galaxies frequently had irregular
optical morphologies (interactions, mergers or compact galaxies), and
higher radio/optical flux ratios than normal spirals, suggesting they
were a population distinct from the brighter blue sources.

Thuan \etal\ (1984)\nocite{Thuan84} observed 48 sources in the LBDS in
the near-infrared $JHK$-bands, to $K\simeq 17.5$~mag.  They discovered
that the optical--infrared colours of the faint radio galaxies were
due to their stellar populations, showing no correlation with their
radio flux or power.  Less than 10\% of the red radio galaxies had
evidence for non-thermal infrared emission.  The colours of the red
galaxies were consistent with those of distant ($z\ga 0.2$)
non-evolving or passively-evolving luminous giant ellipticals, such as
observed in the 3CR survey \cite{Lilly84}.  The optical/infrared
colours of the faint blue radio galaxies were indicative of
low-luminosity and/or star-forming galaxies.

High-resolution maps of many of the LBDS sources were obtained at
1.4~GHz with the Very Large Array (VLA), in order to investigate the
radio morphologies of weak radio sources \cite{Oort87,Oort88a}.  The
median angular size of the radio samples was found to decrease with
decreasing flux density, to $\sim 2$\arcsec\ at 1--10~mJy and $\la
1$\arcsec\ at 0.4--1.0~mJy.  This was attributed to the combined
effects of a change in the population and a decrease in the intrinsic
size of radio sources at lower (radio) luminosity.  Approximately 90\%
of the red galaxies were resolved by the 1\farcs4 beam, but half of
the blue radio galaxies remained unresolved.  There was evidence in
the sample for a decrease in intrinsic size with increasing redshift
-- a factor of 4 from $z=0$ to $z=0.5$.  The morphological information
and radio positions from this survey have been incorporated into
Table~\ref{radiotable}.  Radio observations have also been made of the
LBDS at additional frequencies, for example: 4.86~GHz (6~cm) by
Donnelly, Partridge \& Windhorst (1987)\nocite{Donnelly87}, 327~MHz
(92~cm) by Oort, Steemers \& Windhorst (1988)\nocite{Oort88}.  Deeper
VLA observations were also made of the Lynx-2 field, reaching
5-$\sigma$ limiting fluxes of 0.2~mJy at 1.41~GHz \cite{Windhorst85}
and $\sim 20$~$\mu$Jy at 8.44~GHz \cite{Windhorst93}.

Several sources drawn from the LBDS Hercules sample have also been the
subject of individual investigations.  The most widely-studied of
these is 53W002, which, at $z=2.390$, is the most distant source to
date in the whole LBDS \cite{Windhorst91}.  A total of 67 orbits with
the {\it Hubble Space Telescope (HST)\/} have now been devoted to
53W002 and its companions.  Windhorst, Keel \& Pascarelle
(1998)\nocite{Windhorst98a} used these data to investigate the
alignment effect and evolution of this relatively weak high-redshift
radio galaxy.  Narrow-band images of the field around the galaxy led
Pascarelle et~al.~(1996a,b)\nocite{Pascarelle96a,Pascarelle96b} to
discover a group of emission-line objects at $z\simeq 2.4$ around the
radio source, which they suggested were subgalactic clumps that would
eventually collapse into a few massive galaxies today.

Spectroscopic observations with the Keck Telescope of two of the
reddest sources (53W069 \& 53W091) have enabled detailed models of
their spectra to be used to estimate their ages.  53W091 was found to
have a likely age of 3.5~Gyr at $z=1.552$ \cite{Dunlop96,Spinrad97},
and a best-fitting age of 4.0~Gyr was determined for 53W069 at
$z=1.432$ \cite{Dey97,Dunlop99}.  However, other authors have
suggested that the age of 53W091 is nearer to 1.5--2.0~Gyr
\cite{Bruzual97,Yi00}.  In addition to 53W002, {\it HST\/}
observations of several other LBDS Hercules sources have been
published: 53W044 \& 53W046 \cite{Keel93,Windhorst94} and 53W069 \&
53W091 \cite{Waddington00b}.

\subsection{Sample completeness and source weights}

Before reviewing the radio properties of the LBDS Hercules sample, it
is necessary to discuss the completeness of the sample and the
associated weighting of radio sources.  Given the selection criteria
of the LBDS ($S_{\rm p}/N \ge 5\sigma$, $A(r) \le 5$), the sample will
be complete only for certain ranges in total flux density $S_\nu$,
angular size $\psi$ and the component flux ratio $f$.  Here, we
summarize the discussion of the completeness, that is given in full in
Windhorst \etal\ (1984a)\nocite{Windhorst84a}.

Due to the primary beam attenuation (i.e.\ the sensitivity of the
telescope decreases with radial distance from the pointing centre),
the observed signal-to-noise of a given source will depend on where it
is on the map. For example, a source that lies just above the
5-$\sigma$ cut at the centre of the map would have fallen below the
cut if it happened to lie at the edge, and it would then not be
included in the sample.  Assuming the primary beam is isotropic, then
the correction for this incompleteness is a simple,
geometrically-determined weight in the source counts and optical
identification statistics.  Every radio source is weighted inversely
proportional to the area over which it could have been seen, before it
dropped below the 5-$\sigma$ threshold.

This primary beam weight is 
\begin{equation}
W_{\rm pb} = \left( {{0.464}\over{r_{\rm lim}}} \right)^2
\label{wpb}
\end{equation}
where $r_{\rm lim}$ is the distance from the beam centre (in degrees) at
which the source would fall below the 5-$\sigma$ cut:
\begin{equation}
r_{\rm lim} = {{1}\over{61.191 \times \nu}} \arccos \left[ \left( 
  {{S_{\rm p}}\over{N}} \right) {{A(r_{\rm obs})}\over{5}} 
  \right]^{-{1\over6}}.
\label{wpb1}
\end{equation}
Here, $\nu$ is the frequency in GHz, $S_{\rm p}/N$ is the observed
peak signal-to-noise, and the attenuation ($A$) at the distance
($r_{\rm obs}$) from the beam centre where the source was observed is
given by Windhorst \etal\ (1984a)\nocite{Windhorst84a} as
\begin{equation}
A(r_{\rm obs}) = \cos^{-6} \left( 61.191 \times \nu \times r_{\rm obs} \right).
\label{attenuation}
\end{equation}

A second, more complicated, source of incompleteness is the
resolution/population bias -- a resolved source of a given sky flux
$S_{^{\rm TOT}}^{_{\rm SKY}}$ will drop below the 5-$\sigma$ cut-off
more easily than a point source of the same $S_{^{\rm TOT}}^{_{\rm
SKY}}$.  Thus the completeness depends on the distribution of source
angular sizes and component flux ratios.  The determination of, and
statistical corrections for, this incompleteness were found by
processing Monte Carlo simulations of artificial data using the same
algorithms as the real data.  They are detailed in Windhorst \etal\
(1984a)\nocite{Windhorst84a}, to which one is referred.  Higher
resolution (arcsecond) VLA images have subsequently shown that the
median angular size of radio sources is a function of flux density,
decreasing monotonically from $\sim 20$\arcsec\ at 1~Jy to $\sim
2$\arcsec\ at 1~mJy.  Windhorst, Mathis \& Neuschaefer
(1990)\nocite{Windhorst90} showed that this resulted in an
overestimate of the original resolution correction, which was based on
the lower resolution WSRT data, by a factor of approximately
twenty-five percent.  The revised source weights were computed
following equation~(14) of Windhorst \etal\
(1984a)\nocite{Windhorst84a}, incorporating the revised angular sizes,
as follows:
\begin{equation}
W_{\rm res} = 1 + {{58\arcsec}\over{{\rm HPBW}_\alpha}} \exp 
   \left( {{-2.71 (S_{\rm TOT}/N)}\over{(S_{\rm p}/N)_{\rm cut-off}}} \right).
\label{wres}
\end{equation}
Here, ${\rm HPBW}_\alpha=12\arcsec$ is the half-power beam-width of
the observations, $S_{^{\rm TOT}}/N$ is the total observed
signal-to-noise of the source and $(S_{\rm p}/N)_{\rm cut-off}=5$ is
the selection criterion of the sample.  The statistical errors in this
correction were shown to be a few percent at the $S_{^{\rm TOT}}\simeq
10\sigma$ level, increasing to about 10\% at $S_{^{\rm TOT}}\simeq
5\sigma$ \cite{Windhorst90}.


\begin{figure}
\psfig{file=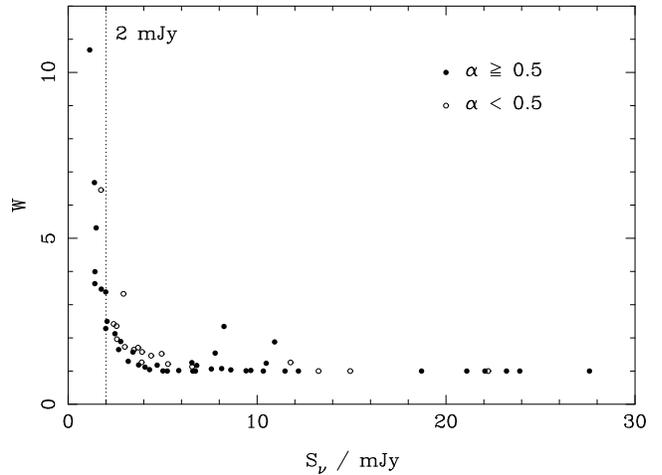,width=85mm}
\caption{Source weights as a function of flux density for the complete
Hercules sample of 72 objects.  Applying the additional selection
criterion $S_\nu \ge 2$~mJy excludes virtually all sources with $W >
2.5$.  Solid and open symbols denote steep-spectrum ($\alpha \ge 0.5$)
and flat-spectrum ($\alpha < 0.5$) sources respectively (where $S_\nu
\propto \nu^{-\alpha}$).\label{hercweights}}
\end{figure}

The total weight of each source is then the multiple of the primary 
beam and resolution weights:
\begin{equation}
W = W_{\rm pb} \times W_{\rm res}.
\label{wtotal}
\end{equation}
These revised weights are given in the last column of
Table~\ref{radiotable}, and are plotted in Fig.~\ref{hercweights}.
This figure clearly shows how the weights rise exponentially near the
survey flux density limit of 1~mJy.  By applying a flux density limit
of 2~mJy to the sample, the ``completeness'' is increased
significantly from 56\% to 76\% -- here defining the ``completeness''
as the ratio of the number of sources actually detected to the total
weighted number of sources.  There are 9 sources with 1~mJy~$\le S_\nu <
2$~mJy that are excluded, with an average weight of 5 (a
``completeness'' of 20\%).  Being the faintest objects, they have the
largest errors and their effect on the sample is greatly exaggerated
by the high weights.  Therefore an a posteriori ``2-mJy sample'' is
defined, consisting of those 63 sources with $S_{1.4} \ge 2$~mJy.
This will used in preference to the full Hercules sample in some of
the statistical studies that follow in this and subsequent papers.  It
must be emphasized that following application of these weights, the
incompleteness of the sample is fully corrected for and so it is, in
effect, {\it complete}.

\subsection{Radio properties of the Hercules sample}

The radio data for the Hercules sample are summarized in
Table~\ref{radiotable}.  The coordinates are from the VLA 1.4~GHz
observations where available \cite{Oort87,Oort88a}, otherwise the WSRT
co-ordinates are used \cite{Windhorst84a}.  The 1.4~GHz and 0.6~GHz
fluxes ($S_{1.4}$ and $S_{0.6}$ respectively) and the resulting
spectral indices ($\alpha^{1.4}_{0.6}$, where $S_\nu \propto
\nu^{-\alpha}$) are from the WSRT data
\cite{Windhorst84,Windhorst84a}.  The radio morphological data are
taken from Oort et~al.\ (1987)\nocite{Oort87} where available,
otherwise the WSRT data are shown.  Column~6 (${\cal R}$) notes
whether the source is unresolved (U), resolved (R) or extended (E).
An unresolved source consists of one component that is not
distinguishably larger than the WSRT beam; a resolved source consists
of one component whose extent is larger than the beam; and a source is
classified as extended if it is clearly resolved into two (the
classical double) or more components.  Column~7 is the largest angular
size ($\psi$) of the source in arcseconds, or an upper limit to $\psi$
for unresolved sources.  The position angle ($\phi$) of the largest
angular size is given in degrees east of north.  The weights ($W$)
have been recalculated as discussed in the previous section and are
given in the final column.


\begin{figure}
\psfig{file=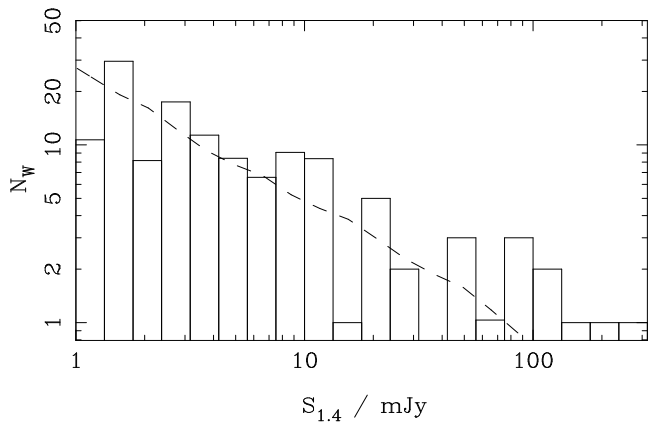,width=85mm}
\caption{The flux density distribution for the Hercules field.  The
histogram is the observed data; the dashed line is the expected
distribution based on the 1.4~GHz source counts (Windhorst \&
Waddington 2000).\label{hercflux}}
\end{figure}

The flux density distribution of all 72 sources is presented as a
histogram in Fig.~\ref{hercflux}.  The median flux is 3.0~mJy, and
for the 2-mJy sample the median is 6.6~mJy.  The dashed line in the
figure is the predicted distribution based on a 5\th-order polynomial
fit to the 1.4~GHz differential source counts of Windhorst \&
Waddington (2000)\nocite{Windhorst98}.  The fit was based on a sample
of 10575 sources drawn from a large number of recent radio surveys.
The overall agreement between data and model is indicative that the
Hercules sample being used here is a fair representation of the full
LBDS and indeed other surveys.  We note that there is a small excess
of bright sources at $S_\nu > 80$~mJy, although, given the small area
of the Hercules field, this may not be significant.


\begin{figure}
\psfig{file=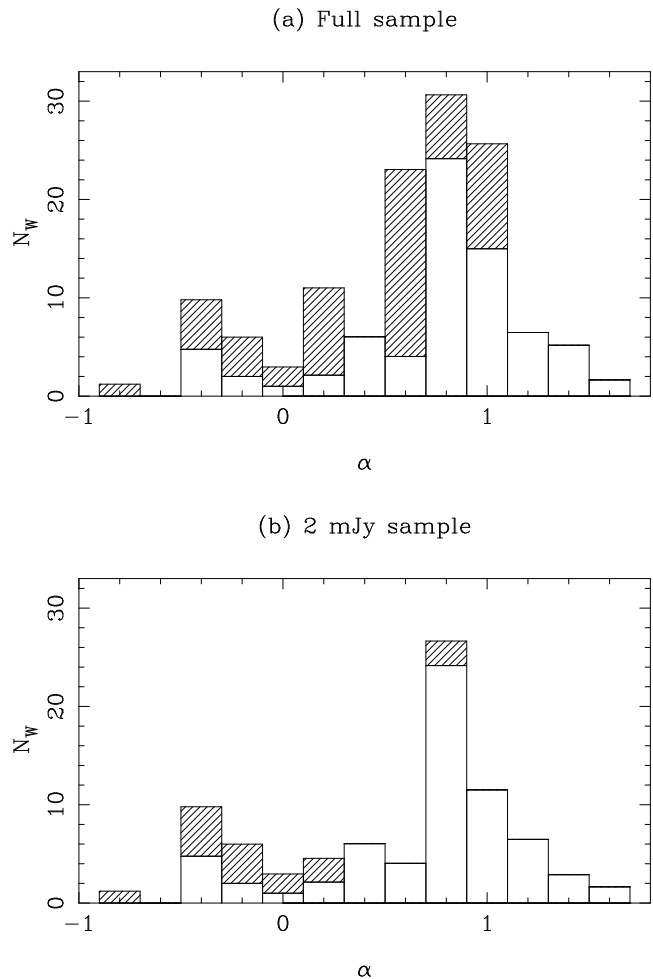,width=85mm}
\caption{The distribution of 1.4--0.6~GHz spectral indices for: (a) the
complete Hercules sample ($S_{1.4} \ge 1$ mJy); and (b) the 2-mJy
sample.  Hatched histograms denote sources that were not detected at
0.6~GHz and thus only have upper limits to $\alpha$.\label{hercalpha}}
\end{figure}

In Fig.~\ref{hercalpha} the distribution of 1.4--0.6~GHz (21--50~cm)
spectral indices is plotted, for both the full Hercules sample and the
2-mJy sample (taking $S_\nu \propto \nu^{-\alpha}$).  Not all sources
were detected at 0.6~GHz and thus those have only upper limits to
their spectral index (indicated by the hatched histogram).  The median
spectral index of those sources with detections at both frequencies
(i.e.\ {\it excluding\/} upper limits) is $\alpha_{\rm med}\simeq 0.8$
for both samples.  Taking into account the 0.6~GHz flux limits gives
lower and upper limits to the median spectral index itself:
$0.05<\alpha_{\rm med}<0.74$ for the full sample, and
$0.74<\alpha_{\rm med}<0.79$ for the 2-mJy sample.  Given the
relatively few sources without spectral indices in the 2-mJy sample,
the true $\alpha_{\rm med}$ is well constrained and we will adopt a
value of 0.77 in subsequent work.  Nearly half the full Hercules
sample is without spectral index information, thus the small lower
limit to $\alpha_{\rm med}$ above.  However, we do not expect the true
value to differ significantly from that of the 2-mJy sample, given
that Donnelly \etal\ (1987)\nocite{Donnelly87} found that the median
spectral index in the Lynx-2 area of the LBDS remains at $\alpha_{\rm
med} \simeq 0.75$ down to $S_{1.4}\simeq 0.25$~mJy, when selected at
1.4~GHz.

\section{Optical identifications and photometry}

Twenty-five of the seventy-two sources in Hercules remained
unidentified on the multi-colour Mayall photographic plates.  These
deep, sky-limited plates reached limiting magnitudes of $U^+ <
23.3$~mag, $J^+ < 23.7$~mag, $F^+ < 22.7$~mag and $N^+ < 21.1$~mag,
roughly corresponding to a giant elliptical galaxy at $z\la 1$
\cite{Windhorst84b,Kron85}.  Thus the unidentified sources were
potentially the most important for investigating the high-redshift
evolution of the faint radio source population.  In order to complete
the survey, deep CCD images were obtained of the unidentified radio
sources.

\subsection{Observations with the Palomar 200-inch}

The Hercules field was observed on the 200-inch Hale telescope at
Palomar Observatory between 1984 and 1988 (Table~\ref{4shobs}), using
the \fsh\ CCD camera \cite{Gunn87}.  This camera consists of four
simultaneously exposed $800\times800$ pixel Texas Instruments CCDs,
covering a nearly contiguous field of $\sim 9\arcmin\times9\arcmin$,
at a scale of 0.332~arcsec~pixel$^{-1}$.  The total field of view is
split by a reflecting pyramid into four separate beams which go to one
each of the detectors.  Poor reflections at the edges of the pyramid
results in a strip of $\approx 10$\arcsec\ of sky between the CCDs
which is not exposed, but $>$95\% of the field is imaged by the
detectors.  One simultaneous exposure of the four CCDs is defined as a
{\it frame\/}.


\begin{table}
\caption{Summary of observing runs\label{4shobs}\label{hercukirt}}
\begin{tabular}{ll}
Dates & Telescope \\
 & \\
1984 July 31 -- August 2   &  Hale 200-inch \\
1985 June 16 -- June 18    &  Hale 200-inch \\
1986 July 31 -- August 3   &  Hale 200-inch \\
1987 May 27 -- May 28      &  Hale 200-inch \\
1987 July 26 -- July 29    &  Hale 200-inch \\
1988 June 7 -- June 8      &  Hale 200-inch \\
1992 July                  &  UKIRT 3.8-m \\
1993 May 15 -- May 17      &  UKIRT 3.8-m \\
1997 June 1 -- June 4      &  UKIRT 3.8-m \\
\end{tabular}
\end{table}

\begin{figure*}
\psfig{file=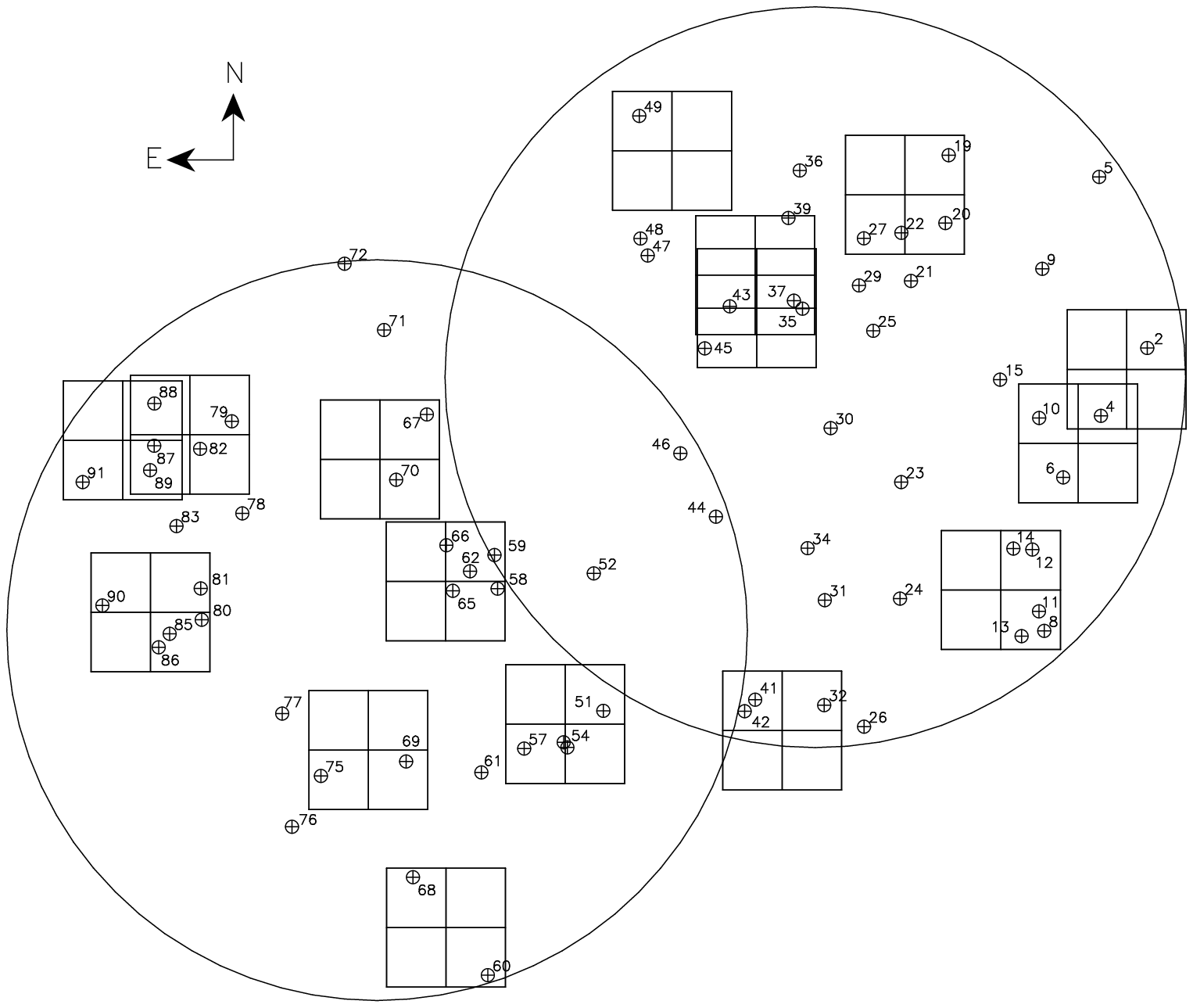,width=160mm}
\caption{Schematic map of the Hercules field.  Each radio source in
the complete sample is denoted by an encircled cross.  The two large
circles show the radial limit of the radio observations (at the $-$7dB
attenuation radius of 0\fdg464).  The boxes show the positions of the
\fsh\ observations.  The centres of the two fields are $17^{\rm h}\,
20^{\rm m}\, 14^{\rm s}$ $+$$49\degr\, 55\arcmin\, 00\arcsec$ and
$17^{\rm h}\, 16^{\rm m}\, 48^{\rm s}$ $+$$50\degr\, 13\arcmin\,
50\arcsec$ (J2000).\label{radiomap}}
\end{figure*}

Sixteen separate pointings of the telescope were used in order to
observe all of the unidentified radio sources (Fig.~\ref{radiomap}).
Multiple observations were made through Gunn $g$, $r$ and $i$ filters
over the six runs.  Photometric standard stars were observed each
night, normally at the start, middle and end of observing.  Many of
the nights were not photometric, though in almost all cases at least
one photometric frame was obtained in each filter for each source.
During the 1984 run, a problem with the \fsh\ dewar resulted in liquid
nitrogen coolant being spilled onto the pyramid thus obscuring much of
the frame.  Although taken in otherwise photometric conditions, these
images were only occasionally used in the analysis, in cases where the
target had not been observed at a later date and where the radio
source was not affected by the spilt liquid nitrogen.  The May 1987
run was used for spectroscopy and also an attempt to observe 53W043 by
occulting a very bright star in the field that obscured the radio
source.  The July 1987 run was primarily spectroscopic data but a few
images were also obtained.


\subsection{Data reduction methods}

The images were processed in a standard manner
\cite{Neuschaefer92,Neuschaefer95a}, but with some important
differences.  Examination of several bias frames confirmed that there
was no two-dimensional structure to the bias (the mean counts in a $10
\times 10$ pixel box varied by $< 0.3$\% across the image).  For
each exposure a first order spline was fitted to the 16-column bias
strip and subtracted from across the image.  The next step is normally
to remove the dark current, however, the \fsh\ CCDs have a very low
dark current, which the observers considered to be negligible.  Tests
on the two best dark frames showed that the counts were comparable to
the noise in the bias frames for exposures of about 1000~s
(corresponding to a dark current of $<
10^{-3}$~e$^-$~s$^{-1}$~pixel$^{-1}$), thus they were not used in the
processing.

Flat-field images of the inside of the telescope dome were made at the
beginning and end of each night.  These images were the average of
eight exposures of 7--10 seconds each for $g$ and 2--4 seconds each
for $r$ and $i$.  For each observing run, the mean of all the
flat-field images was calculated, after scaling and rejection of bad
pixels, to give a single flat-field for the whole of the run.  The
mean value of all the pixels in all four CCDs of the flat-field frame
was calculated, and each image divided by this single value.  Scaling
in this way corrected for the global sensitivity differences between
the four CCDs, assuming that they had been uniformly illuminated.

In some of the $g$ and $r$ images and in almost all the $i$ images,
large-scale gradients remained across the CCDs, following division by
the appropriate flat-field.  Neuschaefer \& Windhorst
(1995)\nocite{Neuschaefer95a} determined that the features in $g$ and
$r$ were due to scattered light rather than sensitivity variations in
the CCDs.  Such global gradients were not important, as variations in
the background of each target were removed at the photometry
stage. The features in the $i$-band images posed a more serious
problem, as they appeared to have two distinct sources.  First, there
were large-scale gradients across the images of a few percent of the
sky background (larger than in $g$ and $r$), similar in appearance to
the flat-field pattern itself.  Rather than being scattered light, it
is suggested that this was due to the different spectral
characteristics of the night sky compared with the lamp used to
illuminate the flat-fields, which was supported by the observation
that the problem was most significant near twilight when the sky was
changing most rapidly.  This was then an extra {\it multiplicative\/}
correction.  The second problem with the $i$ images was that of
fringing due to the strong night sky lines, on a smaller scale than
the large gradients.  This is an {\it additive\/} correction that
scales with the brightness of the sky background.

Following Neuschaefer \& Windhorst (1995)\nocite{Neuschaefer95a}, who
actually only considered the fringing, an $i$-band `superflat' was
constructed.  Most of the $i$-band data were taken in the June 1988
run, in which there were approximately 50 $i$-band exposures.
Observations of the same region of sky were offset by
10\arcsec--20\arcsec\ and repeated no more than three or four times
during the run, thus all the frames were effectively pointing at
different parts of the sky.  Following infrared imaging techniques, a
superflat was constructed from the median of all 50 frames.  This
produced a high signal-to-noise picture of the sky flat-field, devoid
of all sources.  The flattened $i$-band images were divided by the
superflat in order to remove the large-scale features.  The fringing
pattern was also removed by this process, although it is really an
additive error.  If fringing had been the dominant problem, then a
scaled version of the superflat should have been subtracted from the
images.  This was done for several frames and the results compared
with those obtained by dividing by the superflat.  There was no visual
difference between the results of the two methods, nor were they
statistically different (both methods increased the sky noise by
0.5--1.0\%).  For the $i$-band observations in earlier years there
were not enough independent exposures to construct a superflat,
however using the 1988 superflat often improved the images from these
earlier runs.

Finally, the observations from different years were combined to
produce an image of each radio source in the $g$, $r$ and $i$ filters.
The images were registered using three or four unsaturated stars --
most images only required a linear offset, although a 0\fdg7 rotation
was applied to the 1988 data.  The images were then stacked, rejecting
any remaining cosmic rays or bad pixels, in those cases where there
were sufficient exposures to make this possible.

\subsection{Optical identifications}

The astrometry was performed on the $r$-band images in general.
Occasionally the radio source was not visible in $r$, in which case
either the $g$ or $i$ image was used.  The Cambridge APM Guide Star
Catalogue \cite{Lewis97} was used to find 10--20 secondary standards
in each image of interest.  This catalogue is based on digitised scans
of the first epoch Palomar Observatory Sky Survey, which has an
internal accuracy of 0\farcs1--0\farcs2, and is accurate to 0\farcs5
externally.  A six-coefficient linear model was fitted to the \fsh\
images, yielding root-mean-square residuals of $0\farcs5 \pm
0\farcs1$.

The 1.4~GHz radio coordinates from Table~\ref{radiotable} were
converted to pixel coordinates on the \fsh\ images.  The errors in the
radio positions are typically 0\farcs5--1\farcs0.  Combining these in
quadrature with the optical astrometry errors defines a circle about
the predicted coordinates of the source, with a 1-$\sigma$ radius of
0\farcs7--1\farcs2, within which $\sim 68$\% of the optical
counterparts to the radio sources are expected to be found.  Strictly,
the errors in right ascension and declination are not equal and the
search area should be an ellipse \cite{Windhorst84b}, however with an
ellipticity of typically 0.1--0.2 the assumption of symmetrical errors
was adequate for all identifications.  Given the faintness of many of
the sources, an optical identification was considered to be real only
if the source was detected in at least two passbands.  This
requirement had the potential to exclude very red or very blue
objects, but in practice no candidate identification was rejected on
this basis.

A total of 47 sources were observed, of which 25 had not been
identified on the Mayall photographic plates.  Of these sources, new
optical identifications for 19 are proposed here, 3 have been
published elsewhere (53W002, 53W069, 53W091) and only 3 sources remain
unidentified (53W037, 53W043, 53W087).  Of the new identifications,
53W054 which was previously classified as a two-component radio source
has now been reclassified as two independent objects with solid
identifications (53W054A \& 53W054B).  Two sources have new
identifications compared with those in Windhorst \etal\
(1984b)\nocite{Windhorst84b}: 53W051 has a close optical counterpart
where it was previously associated with a source offset along the
radio axis; and 53W022 is identified with a fainter source 5\arcsec\
to the north-west of the original candidate.  The coordinates of all
the identifications are given in Table~\ref{photmaster} and the \fsh\
images are presented in Fig.~\ref{images}.


\begin{table*}
\begin{minipage}{15.0cm}
\caption{LBDS Hercules sample -- optical \& infrared data.\label{photmaster}}
\begin{tabular}{lrrrcccccccl}

Name\ \ \ \ \ \ \ \  & \multispan3{\hfil RA (J2000)\hfil} & $z$ &  $g$   &  $r$   &  $i$  &  $J$  &  $H$  &  $K$  &  References\sn{\P} \\
       & \multispan3{\hfil Dec (J2000)\hfil} & Type\sn{*} & $\Delta g$ & $\Delta r$ & $\Delta i$ & $\Delta J$ & $\Delta H$ & $\Delta K$ & \\
 & & & & & & & & & & & \\

53W002 & 17 & 14 & 14.70 & 2.390 & 23.42  & 23.26  & 23.13  & 20.71  & 20.04  & 18.87  & W91, W98 \\
       & 50 & 15 & 29.9  &   G   & (0.12) & (0.09) & (0.10) & (0.16) & (0.17) & (0.12) & \\
53W004 & 17 & 14 & 36.62 &   -   & 24.38  & 24.05  & 23.25  &   -    & 19.78  & 19.27  & \\
       & 50 & 10 & 25.4  &   G   & (0.22) & (0.16) & (0.11) &   -    & (0.16) & (0.27) & \\
53W005\sn{\dagger} & 17 & 14 & 36.73 & 0.95 & 23.34  & 23.02  & 21.73  &   -    &   -    & 16.97  & II, T84 \\
       & 50 & 28 & 23.1  &   G   & (0.30) & (0.30) & (0.30) &   -    &   -    & (0.25) & \\
53W008 & 17 & 15 &  3.90 & 0.733 & 20.61  & 20.31  & 20.02  & 17.68  &   -    & 16.57  & B82, T84 \\
       & 49 & 54 & 18.2  &   Q   & (0.06) & (0.04) & (0.05) & (0.28) &   -    & (0.21) & \\
53W009\sn{\dagger} & 17 & 15 &  3.56 & 1.090 & 17.57  & 17.72  & 18.30  & 16.42  & 16.12  & 15.15  & K85, T84 \\
       & 50 & 21 & 30.8  &   Q   & (0.30) & (0.30) & (0.30) & (0.10) & (0.11) & (0.11) & \\
53W010\sn{\ddagger} & 17 & 15 &  5.35 & 0.48  & 21.08  & 19.35  & 19.33  & 17.10  & 16.38  & 15.47  & II, T84 \\
       & 50 & 10 & 19.7  &   G   & (0.30) & (0.05) & (0.30) & (0.14) & (0.15) & (0.07) & \\
53W011 & 17 & 15 &  6.06 &   -   & 21.45  & 20.61  & 20.29  & 18.27  & 17.54  & 17.10  & NK \\
       & 49 & 55 & 45.6  &   G   & (0.06) & (0.04) & (0.05) & (0.15) & (0.12) & (0.15) & \\
53W012 & 17 & 15 &  9.10 & 1.328 & 23.97  & 23.98  & 23.41  &   -    & 20.75  & 19.25  & II \\
       & 50 &  0 & 25.5  &   G   & (0.11) & (0.11) & (0.10) &   -    & (0.39) & (0.40) & \\
53W013 & 17 & 15 & 14.36 &   -   & 24.79  & 24.77  & 25.12  &   -    & \llap{$>$}21.1 & 20.37  & \\
       & 49 & 53 & 56.7  &   G   & (0.18) & (0.19) & (0.33) &   -    &   -    & (0.51) & \\
53W014 & 17 & 15 & 17.79 &   -   & 23.55  & 23.16  & 22.83  &   -    &   -    & 18.46  & \\
       & 50 &  0 & 29.7  &   Q   & (0.08) & (0.06) & (0.07) &   -    &   -    & (0.16) & \\
53W015\sn{\dagger} & 17 & 15 & 23.64 & 1.129 & 19.41  & 18.91  & 19.37  &   -    &   -    & 16.18  & II, NK \\
       & 50 & 13 & 13.1  &   Q   & (0.30) & (0.30) & (0.30) &   -    &   -    & (0.16) & \\
53W019\sn{\ddagger} & 17 & 15 & 46.84 & 0.542 & 22.40  & 21.22  & 20.91  &   -    &   -    & 16.57  & II, NK \\
       & 50 & 30 &  6.3  &   Q?  & (0.30) & (0.06) & (0.30) &   -    &   -    & (0.25) & \\
53W020\sn{\ddagger} & 17 & 15 & 48.63 & 0.100 & 17.73  & 16.86  & 17.01  & 15.36  & 14.60  & 14.21  & K85, T84 \\
       & 50 & 24 & 59.7  &   G   & (0.30) & (0.05) & (0.30) & (0.07) & (0.06) & (0.06) & \\
53W021\sn{\dagger} & 17 & 16 &  4.75 &   -   &   -    & 22.56  & 21.51  &   -    & 18.38  & 16.37  & T84 \\
       & 50 & 20 & 42.2  &   G   &   -    & (0.30) & (0.30) &   -    & (0.45) & (0.15) & \\
53W022\sn{\ddagger} & 17 & 16 &  8.99 & 0.528 & 21.84  & 21.32  & 20.64  &   -    &   -    & 16.78  & II \\
       & 50 & 24 & 17.3  &   G   & (0.30) & (0.06) & (0.30) &   -    &   -    & (0.06) & \\
53W023\sn{\dagger} & 17 & 16 & 10.32 & 0.57 & 22.12  & 21.00  & 20.08  & 17.28  & 16.71  & 15.85  & II, NK \\
       & 50 &  5 & 33.4  &   G   & (0.30) & (0.30) & (0.30) & (0.17) & (0.14) & (0.08) & \\
53W024\sn{\dagger} & 17 & 16 & 11.19 & 1.961 & 20.65  & 21.01  & 20.64  &   -    &   -    & 16.77  & H96, NK \\
       & 49 & 56 & 48.7  &   Q   & (0.30) & (0.30) & (0.30) &   -    &   -    & (0.32) & \\
53W025\sn{\dagger} & 17 & 16 & 22.65 &   -   & 23.10  & 22.90  & 21.74  &   -    &   -    & 18.09  & NK \\
       & 50 & 16 & 59.1  &   ?   & (0.30) & (0.30) & (0.30) &   -    &   -    & (0.60) & \\
53W026\sn{\dagger} & 17 & 16 & 28.27 & 0.55  & 22.23  & 21.44  & 20.70  &   -    &   -    & 16.39  & II, T84 \\
       & 49 & 47 & 12.2  &   G   & (0.30) & (0.30) & (0.30) &   -    &   -    & (0.21) & \\
53W027\sn{\ddagger} & 17 & 16 & 26.52 & 0.403 & 22.56  & 22.63  & 21.66  &   -    &   -    & 19.74  & II \\
       & 50 & 23 & 53.5  &   G?  & (0.30) & (0.09) & (0.30) &   -    &   -    & (0.38) & \\
53W029\sn{\dagger} & 17 & 16 & 29.24 &   -   & 23.06  & 22.06  & 22.09  &   -    &   -    & 16.86  & NK \\
       & 50 & 20 & 22.0  &   Q   & (0.30) & (0.30) & (0.30) &   -    &   -    & (0.32) & \\
53W030\sn{\dagger} & 17 & 16 & 42.98 & 0.35 & 21.07  & 20.04  & 19.61  & 16.74  & 16.06  & 15.54  & II, NK \\
       & 50 &  9 & 40.1  &   G   & (0.30) & (0.30) & (0.30) & (0.16) & (0.11) & (0.11) & \\
53W031\sn{\dagger} & 17 & 16 & 46.39 & 0.628 & 22.21  & 21.09  & 19.95  & 17.28  & 16.46  & 15.84  & K85, NK \\
       & 49 & 56 & 43.6  &   G   & (0.30) & (0.30) & (0.30) & (0.08) & (0.07) & (0.04) & \\
53W032 & 17 & 16 & 47.15 & 0.37  & 19.51  & 18.48  & 18.17  & 16.60  & 15.87  & 15.21  & II, NK \\
       & 49 & 48 & 46.9  &   G   & (0.07) & (0.07) & (0.04) & (0.12) & (0.10) & (0.14) & \\
53W034\sn{\dagger} & 17 & 16 & 54.45 & 0.281 & 22.38  & 22.18  & 22.87  &   -    &   -    & 19.50  & II \\
       & 50 &  0 & 34.4  &   G   & (0.30) & (0.30) & (0.30) &   -    &   -    & (0.25) & \\
53W035 & 17 & 16 & 55.75 &   -   & 23.74  & 23.66  & 23.39  &   -    & 20.03  & 19.04  & \\
       & 50 & 18 & 38.8  &   ?   & (0.10) & (0.09) & (0.13) &   -    & (0.20) & (0.25) & \\
53W036\sn{\dagger} & 17 & 16 & 56.56 &   -   & 21.69  & 21.86  & 21.76  &   -    &   -    & 19.05  & \\
       & 50 & 29 &  3.4  &   Q?  & (0.30) & (0.30) & (0.30) &   -    &   -    & (0.20) & \\
53W037 & \n & \n & \n    &   -   &   -    & \llap{$>$}25.0 &   -    &   -    &   -    & \llap{$>$}20.7 & \\
       & \n & \n & \n    &   -   &   -    &   -          &   -    &   -    &   -    &   -    & \\
53W039 & 17 & 17 &  1.60 & 0.402 & 20.94  & 19.26  & 18.75  & 16.54  & 15.71  & 15.01  & K85, NK \\
       & 50 & 25 & 28.8  &   G   & (0.09) & (0.05) & (0.03) & (0.11) & (0.10) & (0.06) & \\
53W041 & 17 & 17 & 19.02 &   -   & 26.31  & 25.07  & 26.26  &   -    &   -    & \llap{$>$}20.8 & \\
       & 49 & 49 & 17.6  &   ?   & (0.43) & (0.23) & (0.91) &   -    &   -    &   -    & \\
53W042 & 17 & 17 & 24.07 &   -   & 24.80  & 25.68  & 25.11  &   -    &   -    & \llap{$>$}20.8 & \\
       & 49 & 48 & 22.9  &   ?   & (0.17) & (0.47) & (0.37) &   -    &   -    &   -    & \\

\end{tabular}
\end{minipage}
\end{table*}

\begin{table*}
\begin{minipage}{15.0cm}
\contcaption{}
\begin{tabular}{lrrrcccccccl}

Name\ \ \ \ \ \ \ \  & \multispan3{\hfil RA (J2000)\hfil} & $z$ &  $g$   &  $r$   &  $i$  &  $J$  &  $H$  &  $K$  &  References\sn{\P} \\
       & \multispan3{\hfil Dec (J2000)\hfil} & Type\sn{*} & $\Delta g$ & $\Delta r$ & $\Delta i$ & $\Delta J$ & $\Delta H$ & $\Delta K$ & \\
 & & & & & & & & & & & \\

53W043 & \n & \n & \n    &   -   &   -    & \llap{$>$}23.0 &   -    &   -    &   -    &   -    & \\
       & \n & \n & \n    &   -   &   -    &   -          &   -    &   -    &   -    &   -    & \\
53W044\sn{\dagger} & 17 & 17 & 36.88 & 0.311 & 19.93  & 18.99  & 18.57  & 16.40  & 15.28  & 14.63  & W94 \\
       & 50 &  3 &  4.6  &   G   & (0.30) & (0.30) & (0.30) & (0.19) & (0.07) & (0.08) & \\
53W045 & 17 & 17 & 41.51 & 0.30  & 20.49  & 19.29  & 18.79  & 16.63  & 15.92  & 15.08  & II, NK \\
       & 50 & 15 & 42.8  &   G   & (0.06) & (0.03) & (0.03) & (0.12) & (0.11) & (0.11) & \\
53W046\sn{\dagger} & 17 & 17 & 53.37 & 0.528 & 21.39  & 20.34  & 19.69  & 17.27  & 16.46  & 15.94  & W94 \\
       & 50 &  7 & 51.8  &   G   & (0.30) & (0.30) & (0.30) & (0.10) & (0.08) & (0.12) & \\
53W047\sn{\dagger} & 17 & 18 &  7.89 & 0.534 & 21.87  & 20.73  & 20.24  & 17.31  & 16.56  & 15.76  & II, NK \\
       & 50 & 22 & 44.9  &   G   & (0.30) & (0.30) & (0.30) & (0.12) & (0.10) & (0.12) & \\
53W048\sn{\dagger} & 17 & 18 & 11.18 & 0.676 & 22.38  & 21.01  & 20.16  & 17.56  & 16.55  & 15.68  & II, NK \\
       & 50 & 24 &  1.4  &   G   & (0.30) & (0.30) & (0.30) & (0.12) & (0.07) & (0.06) & \\
53W049 & 17 & 18 & 11.29 & 0.23  & 21.86  & 20.70  & 19.99  & 17.85  & 16.75  & 16.15  & II, NK \\
       & 50 & 33 & 12.7  &   G   & (0.09) & (0.03) & (0.03) & (0.15) & (0.12) & (0.15) & \\
53W051 & 17 & 18 & 30.28 &   -   & 24.55  & 22.84  & 21.87  &   -    &   -    & 17.23  & \\
       & 49 & 48 & 30.2  &   G?  & (0.11) & (0.05) & (0.04) &   -    &   -    & (0.07) & \\
53W052\sn{\dagger} & 17 & 18 & 34.14 & 0.46 & 22.37  & 21.20  & 20.63  &   -    & 17.85  & 16.90  & II, T84 \\
       & 49 & 58 & 50.9  &   G   & (0.30) & (0.30) & (0.30) &   -    & (0.32) & (0.16) & \\
53W054A& 17 & 18 & 47.32 &   -   & 24.52  & 23.93  & 23.30  &   -    & 19.52  & 18.34  & \\
       & 49 & 45 & 48.4  &   G   & (0.26) & (0.16) & (0.13) &   -    & (0.14) & (0.13) & \\
53W054B& 17 & 18 & 50.10 &   -   & 24.26  & 24.04  & 23.59  &   -    & 20.79  & \llap{$>$}19.8 & \\
       & 49 & 46 & 14.5  &   ?   & (0.18) & (0.15) & (0.17) &   -    & (0.34) &   -    & \\
53W057 & 17 & 19 &  9.87 &   -   & 24.75  & 25.29  & 26.14  &   -    & \llap{$>$}21.9 & 21.22  & \\
       & 48 & 45 & 44.5  &   G   & (0.15) & (0.21) & (0.59) &   -    &   -    & (0.51) & \\
53W058 & 17 & 19 & 18.71 & 0.034 & 15.92  & 15.42  & 15.05  &   -    &   -    & 11.92  & K85 \\
       & 49 & 57 & 41.0  &   G   & (0.09) & (0.05) & (0.04) &   -    &   -    & (0.01) & \\
53W059 & 17 & 19 & 20.33 &   -   & 24.53  & 24.59  & 23.92  &   -    & 20.09  & 19.20  & \\
       & 50 &  0 & 20.3  &   G?  & (0.17) & (0.18) & (0.18) &   -    & (0.20) & (0.19) & \\
53W060 & 17 & 19 & 25.31 &   -   & 26.52  & 25.52  & 25.76  &   -    &   -    & \llap{$>$}20.8 & \\
       & 49 & 28 & 45.8  &   G   & (0.53) & (0.26) & (0.45) &   -    &   -    &   -    & \\
53W061\sn{\dagger} & 17 & 19 & 27.40 &   -   & 21.95  & 21.26  & 20.81  &   -    &   -    & 17.39  & T84 \\
       & 49 & 43 & 59.6  &   Q?  & (0.30) & (0.30) & (0.30) &   -    &   -    & (0.26) & \\
53W062 & 17 & 19 & 31.92 & 0.61  & 22.54  & 21.25  & 20.54  &   -    & 17.13  & 17.09  & II, NK, T84 \\
       & 49 & 59 &  6.2  &   G   & (0.11) & (0.06) & (0.05) &   -    & (0.20) & (0.25) & \\
53W065 & 17 & 19 & 39.78 & 1.185 & 23.26  & 22.92  & 22.52  &   -    &   -    & 18.80  & II \\
       & 49 & 57 & 39.3  &   G   & (0.08) & (0.06) & (0.05) &   -    &   -    & (0.21) & \\
53W066 & 17 & 19 & 42.95 &   -   & 24.59  & 24.60  & 24.60  &   -    & 21.34  & \llap{$>$}19.3 & \\
       & 50 &  1 &  3.1  &   G?  & (0.18) & (0.20) & (0.32) &   -    & (0.32) &   -    & \\
53W067 & 17 & 19 & 51.56 & 0.759 & 24.06  & 21.76  & 21.70  &   -    & 18.38  & 18.95  & II \\
       & 50 & 10 & 55.6  &   G   & (0.13) & (0.07) & (0.05) &   -    & (0.06) & (0.20) & \\
53W068 & 17 & 19 & 59.66 &   -   & 22.66  & 22.54  & 22.65  &   -    &   -    & 19.36  & \\
       & 49 & 36 &  9.3  &   ?   & (0.08) & (0.05) & (0.05) &   -    &   -    & (0.28) & \\
53W069 & 17 & 20 &  2.52 & 1.432 & 26.49  & 25.04  & 24.35  & 20.25  & 19.75  & 18.53  & D97, D99\\
       & 49 & 44 & 51.0  &   G   & (0.60) & (0.22) & (0.14) & (0.14) & (0.18) & (0.11) & \\
53W070 & 17 & 20 &  6.07 &   -   & 24.55  & 22.34  & 21.81  &   -    &   -    & 16.89  & \\
       & 50 &  6 &  1.7  &   G   & (0.20) & (0.10) & (0.05) &   -    &   -    & (0.07) & \\
53W071\sn{\dagger} & 17 & 20 & 11.42 & 0.287 & 21.46  & 21.09  & 20.94  &   -    & 18.83  &   -    & K85, NK \\
       & 50 & 17 & 17.4  &   G   & (0.30) & (0.30) & (0.30) &   -    & (0.26) &   -    & \\
53W072\sn{\dagger} & 17 & 20 & 29.54 & 0.019 & 15.71  & 15.27  & 15.04  &   -    &   -    &   -    & K85 \\
       & 50 & 22 & 38.0  &   G   & (0.30) & (0.30) & (0.30) &   -    &   -    &   -    & \\
53W075 & 17 & 20 & 42.34 & 2.150 & 22.16  & 21.35  & 20.73  &   -    & 18.14  & 16.74  & K85, T84 \\
       & 49 & 43 & 48.7  &   Q   & (0.08) & (0.05) & (0.04) &   -    & (0.37) & (0.23) & \\
53W076\sn{\dagger} & 17 & 20 & 55.96 & 0.390 & 20.85  & 19.47  & 19.03  & 16.77  & 16.13  & 15.56  & K85, T84 \\
       & 49 & 40 &  2.8  &   G   & (0.30) & (0.30) & (0.30) & (0.11) & (0.11) & (0.10) & \\
53W077\sn{\dagger} & 17 & 21 &  1.34 & 0.80 & 22.92  & 21.47  & 20.68  &   -    & 17.86  &   -    & II, NK \\
       & 49 & 48 & 34.4  &   G   & (0.30) & (0.30) & (0.30) &   -    & (0.20) &   -    & \\
53W078\sn{\dagger} & 17 & 21 & 18.22 & 0.27 & 20.00  & 19.18  & 18.59  & 16.25  & 15.55  & 14.91  & II, T84 \\
       & 50 &  3 & 35.3  &   G   & (0.30) & (0.30) & (0.30) & (0.09) & (0.11) & (0.06) & \\
53W079 & 17 & 21 & 22.84 & 0.548 & 21.79  & 20.23  & 19.67  & 17.56  & 16.67  & 15.86  & K85, NK \\
       & 50 & 10 & 30.2  &   G   & (0.07) & (0.04) & (0.03) & (0.09) & (0.09) & (0.09) & \\
54W080 & 17 & 21 & 37.50 & 0.546 & 18.29  & 18.41  & 18.09  & 16.72  & 15.84  & 15.02  & K85, T84 \\
       & 49 & 55 & 36.6  &   G   & (0.12) & (0.10) & (0.04) & (0.06) & (0.10) & (0.09) & \\

\end{tabular}
\end{minipage}
\end{table*}

\begin{table*}
\begin{minipage}{15.0cm}
\contcaption{}
\begin{tabular}{lrrrcccccccl}

Name\ \ \ \ \ \ \ \  & \multispan3{\hfil RA (J2000)\hfil} & $z$ &  $g$   &  $r$   &  $i$  &  $J$  &  $H$  &  $K$  &  References\sn{\P} \\
       & \multispan3{\hfil Dec (J2000)\hfil} & Type\sn{*} & $\Delta g$ & $\Delta r$ & $\Delta i$ & $\Delta J$ & $\Delta H$ & $\Delta K$ & \\
 & & & & & & & & & & & \\

53W081 & 17 & 21 & 37.87 & 2.060 & 24.84  & 24.64  & 24.57  &   -    &   -    & 18.91  & S97a \\
       & 49 & 57 & 57.1  &   ?   & (0.20) & (0.23) & (0.29) &   -    &   -    & (0.21) & \\
53W082 & 17 & 21 & 37.62 &   -   & 26.92  & 25.79  & 25.43  &   -    & 21.73  & 21.36  & \\
       & 50 &  8 & 27.4  &   G   & (0.60) & (0.38) & (0.31) &   -    & (0.34) & (0.68) & \\
53W083\sn{\dagger} & 17 & 21 & 48.92 & 0.628 & 22.60  & 22.20  & 21.01  &   -    &   -    & 17.34  & II \\
       & 50 &  2 & 40.1  &   G   & (0.30) & (0.30) & (0.30) &   -    &   -    & (0.10) & \\
53W085 & 17 & 21 & 52.54 & 1.35  & 22.57  & 22.59  & 22.24  & 18.40  &   -    & 16.87  & II, NK, T84 \\
       & 49 & 54 & 33.7  &   Q   & (0.11) & (0.11) & (0.05) & (0.32) &   -    & (0.16) & \\
53W086 & 17 & 21 & 57.67 & 0.40  & 23.91  & 22.36  & 21.25  &   -    & 17.29  & 16.41  & II, T84 \\
       & 49 & 53 & 33.7  &   G   & (0.13) & (0.11) & (0.04) &   -    & (0.21) & (0.10) & \\
53W087 & \n & \n & \n    &   -   &   -    & \llap{$>$}25.0 &   -    &   -    &   -    & \llap{$>$}19.3 & \\
       & \n & \n & \n    &   -   &   -    &   -    &   -    &   -    &   -    &   -    & \\
53W088 & 17 & 21 & 58.92 & 1.773 & 24.46  & 24.30  & 24.03  &   -    &   -    & 19.83  & S97a \\
       & 50 & 11 & 52.4  &   G?  & (0.12) & (0.13) & (0.13) &   -    &   -    & (0.29) & \\
53W089 & 17 & 22 &  1.02 & 0.635 & 25.09  & 24.70  & 24.41  &   -    &   -    & \llap{$>$}19.2 & II \\
       & 50 &  6 & 51.7  &   ?   & (0.21) & (0.19) & (0.33) &   -    &   -    &   -    & \\
53W090 & 17 & 22 & 24.04 & 0.094 & 17.68  & 17.13  & 16.61  &   -    &   -    & 13.84  & K85 \\
       & 49 & 56 & 42.7  &   G   & (0.12) & (0.11) & (0.04) &   -    &   -    & (0.02) & \\
53W091 & 17 & 22 & 32.71 & 1.552 & 25.99  & 25.88  & 24.37  & 20.5   & 19.5   & 18.67  & D96, S97b \\
       & 50 &  6 &  1.4  &   G   & (0.49) & (0.48) & (0.19) & (0.1)  & (0.1)  & (0.13) & \\

\end{tabular}

\medskip

$*$ The object classification is based on the optical appearance, and
spectral features when available.  The sources are divided into
``galaxies'' and ``quasars'' where: G (G?) is a clearly (most likely)
extended source; Q (Q?) is a (most likely) stellar object; and sources
denoted as ? are too faint to classify.

$\dagger$ Coordinates are taken from Windhorst \etal\
(1984b)\nocite{Windhorst84b} and converted to a J2000 equinox.
Photometry is taken from Kron \etal\ (1985)\nocite{Kron85} and
transformed from $J^+F^+N^+$ data using
equations~\ref{gritrans}--\ref{gritrans1}.

$\ddagger$ $g$ \& $i$ photometry is taken from Kron \etal\
(1985)\nocite{Kron85} and transformed from $J^+F^+N^+$ data using
equations~\ref{gritrans}--\ref{gritrans1}.

$\P$ Spectroscopy and infrared photometry references: Butcher, private
communication (1982; B82); Dey (1997; D97)\nocite{Dey97}; Dunlop
\etal\ (1996; D96)\nocite{Dunlop96}; Dunlop (1999;
D99)\nocite{Dunlop99}; Hall \etal\ (1996; H96)\nocite{Hall96}; Kron
\etal\ (1985; K85)\nocite{Kron85}; G. Neugebauer, P. Katgert \etal,
private communications (NK); Spinrad, private communication (1997;
S97a); Spinrad \etal\ (1997; S97b)\nocite{Spinrad97}; Thuan \etal\
(1984; T84)\nocite{Thuan84}; Waddington (1998;
W98)\nocite{Waddington98}; Waddington \etal\ (2000a; paper II);
Windhorst \etal\ (1991; W91)\nocite{Windhorst91}; Windhorst \etal\
(1994; W94)\nocite{Windhorst94}.

\end{minipage}
\end{table*}

Three radio sources were not identified either on the photographic
plates or in the \fsh\ CCD observations, and a few comments can be
made concerning their nature.  The first of these, 53W043, is located
5\arcsec\ from a $R=14$~mag star, and lies within the wings of the
star's point spread function (PSF) in the \fsh\ data.  The PSF of the
photographic plates is smaller than that of the CCDs, giving an upper
limit to the brightness of the optical counterpart of $F^+ > 23$~mag.
Counterparts to 53W037 and 53W087 were not identified on good \fsh\
images, even after both smoothing the images and stacking the $g$, $r$
\& $i$ images together.  For both sources, an upper limit of $r >
25$~mag was calculated.  Neither source was detected in the infrared
(see \S4 below), with limits of $K>20.7$~mag for 53W037 and
$K>19.3$~mag for 53W087.  Both these objects are extended,
steep-spectrum radio sources, suggesting that they may be classical
radio galaxies.  If so, then the faintness of their host galaxies
suggests that they are either at very high redshift, or are obscured
by dust, or are dusty high-redshift sources \cite{Waddington99}.


\begin{figure*}
\begin{minipage}{15cm}

\psfig{file=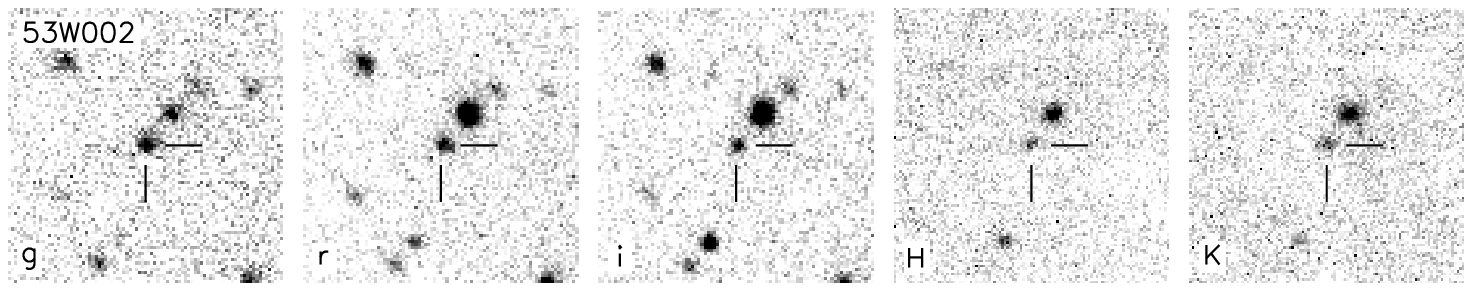}
\psfig{file=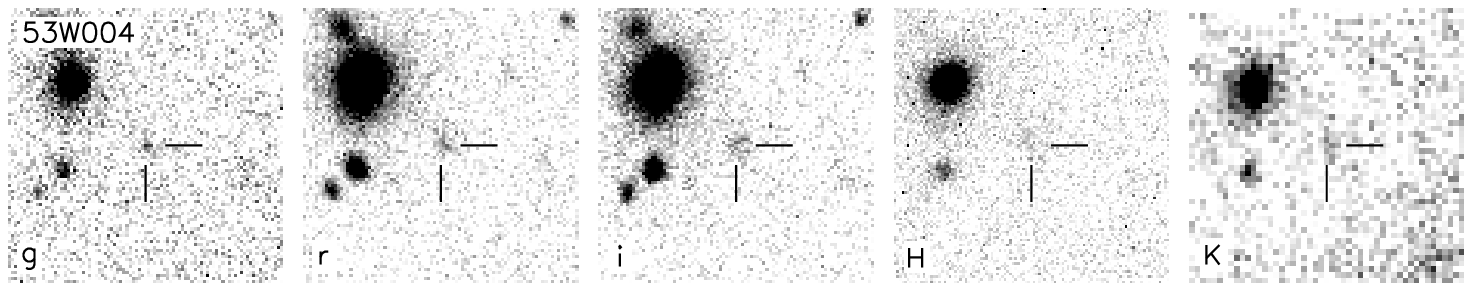}
\psfig{file=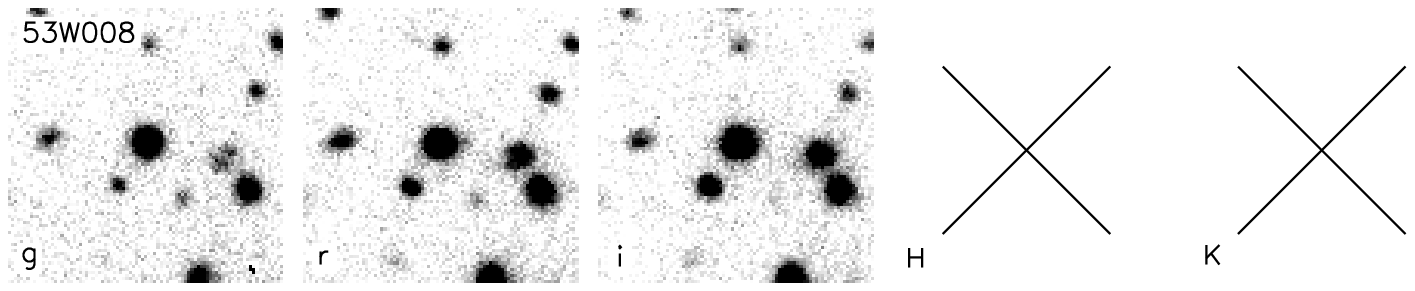}
\psfig{file=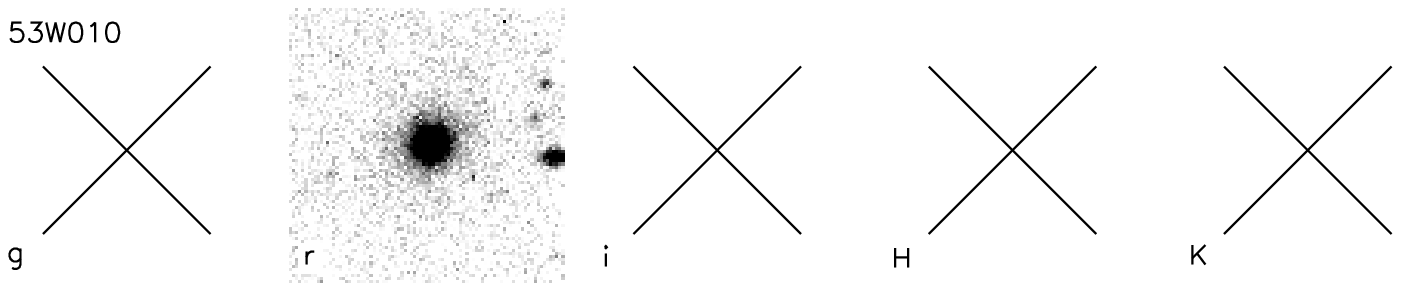}
\psfig{file=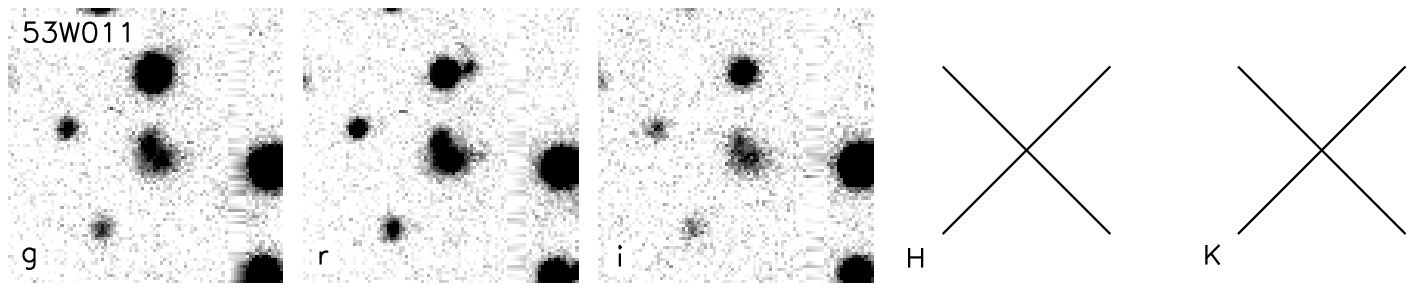}
\psfig{file=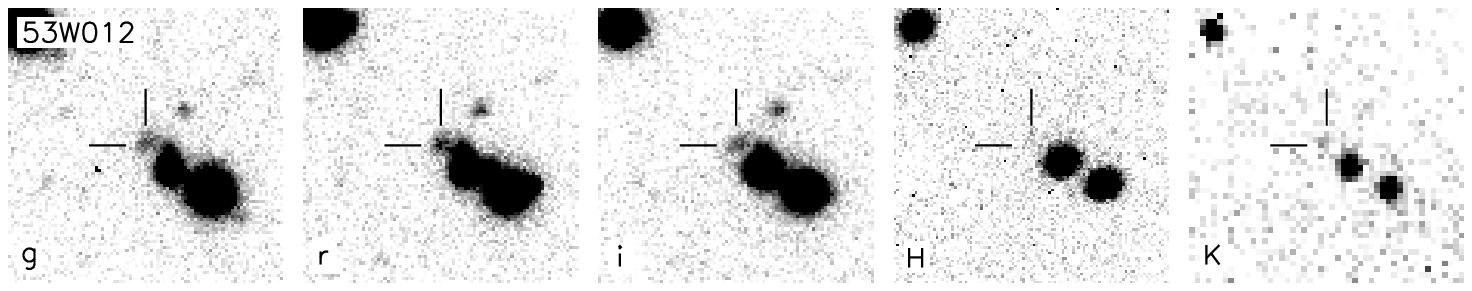}
\psfig{file=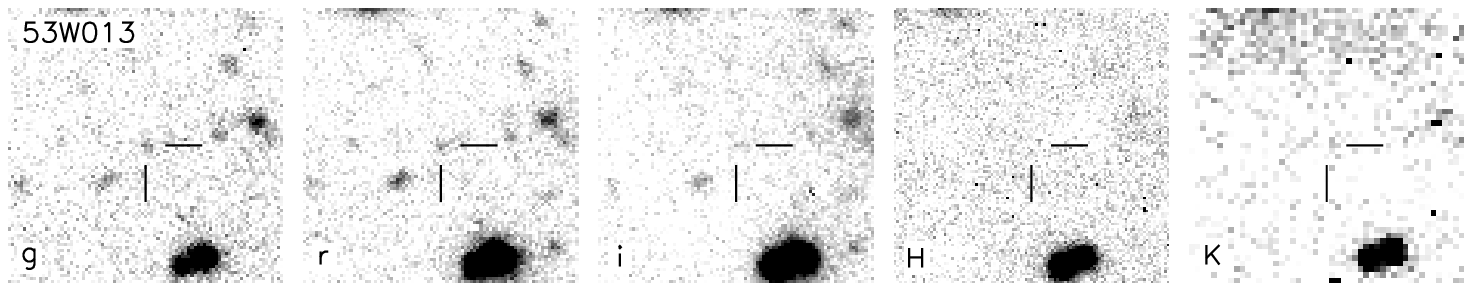}

\caption{$griHK$ images (from left to right) for those sources in the
LBDS Hercules sample that were identified on the optical CCD images
discussed in section~3.  Only three of these sources (53W002, 53W069,
\& 53W091) had $J$-band observations which have been published
elsewhere.  Each image is 30~arcsec on a side, and is orientated with
north to the top, east to the left.  Crosshairs denote the radio
source where necessary.\label{images}}
\end{minipage}
\end{figure*}

\begin{figure*}
\begin{minipage}{15cm}

\psfig{file=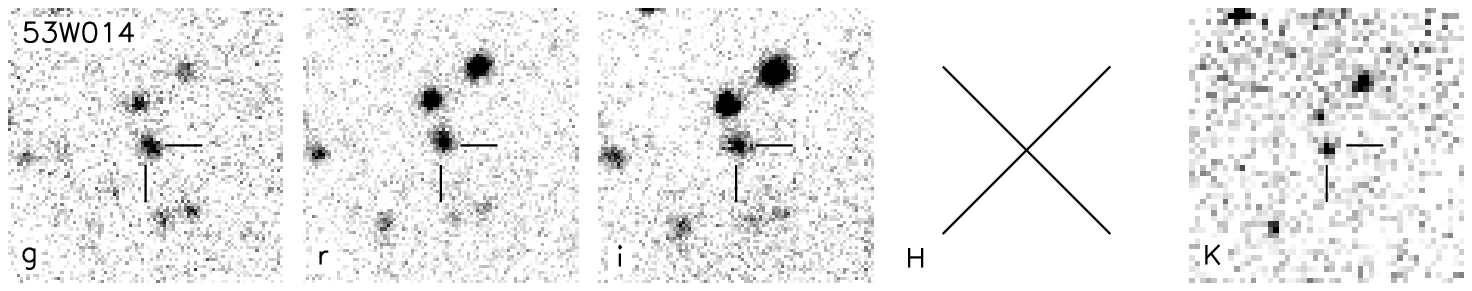}
\psfig{file=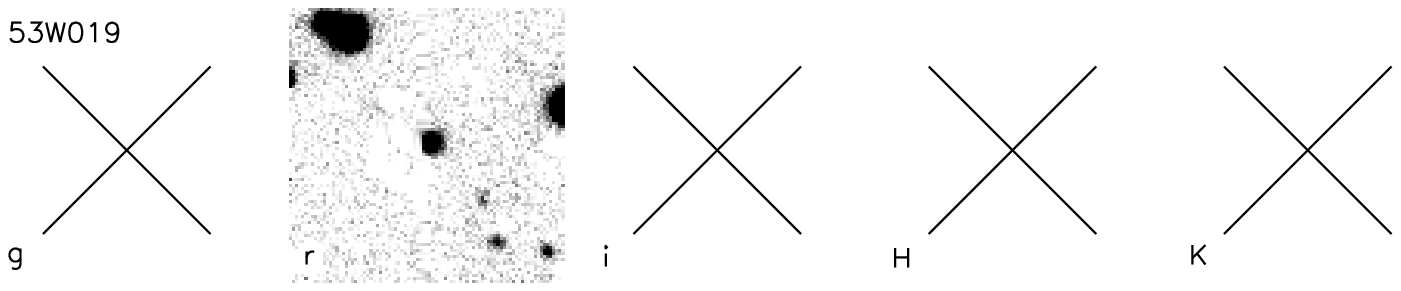}
\psfig{file=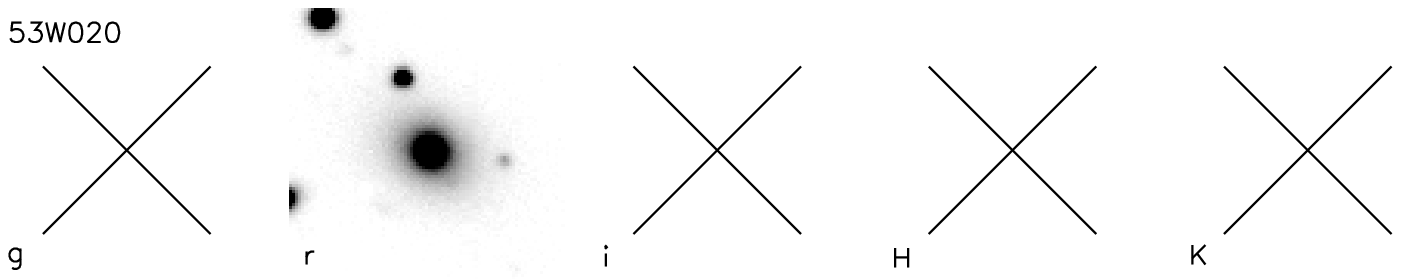}
\psfig{file=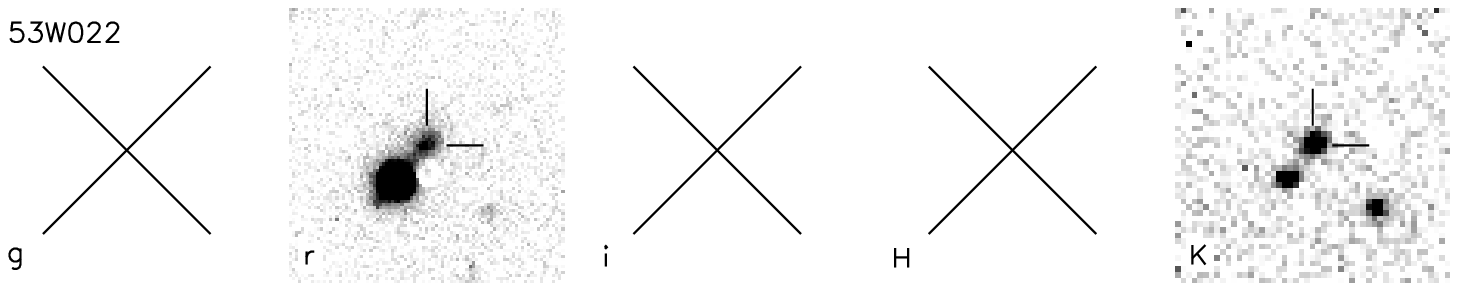}
\psfig{file=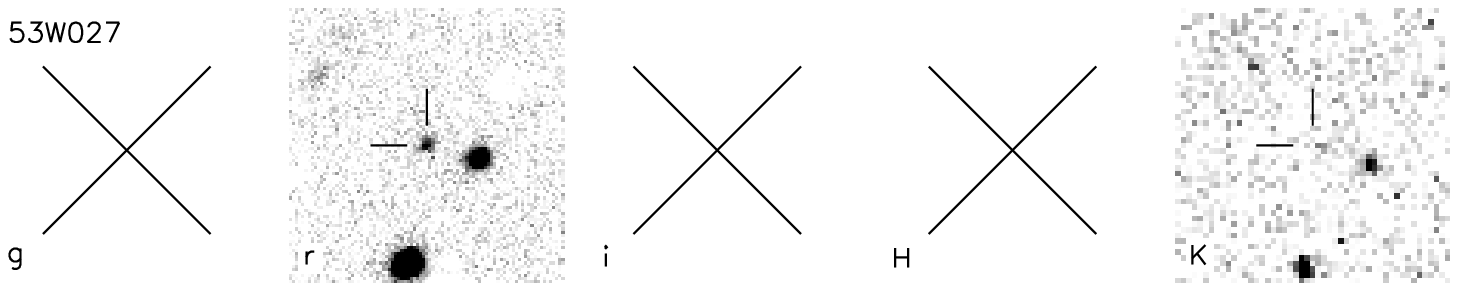}
\psfig{file=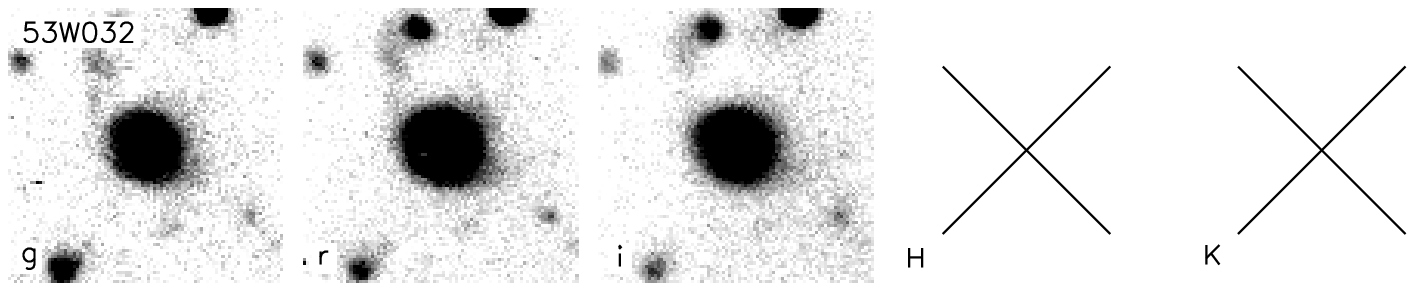}
\psfig{file=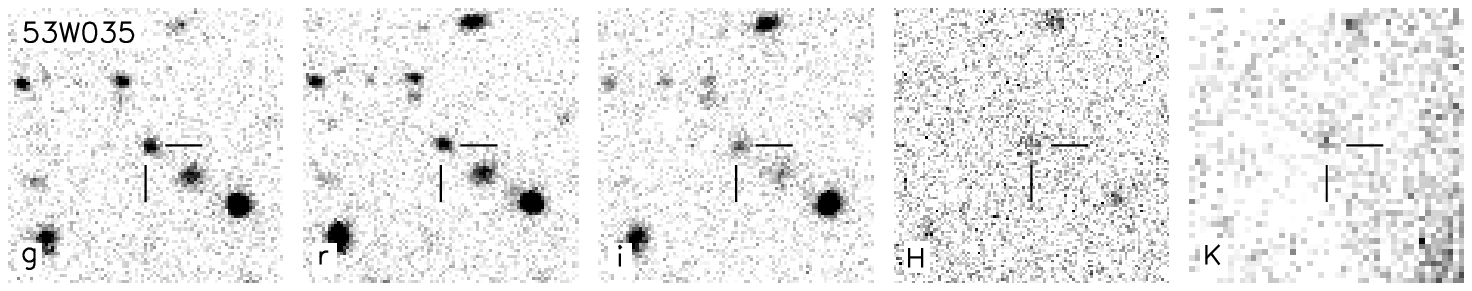}

\contcaption{}
\end{minipage}
\end{figure*}

\begin{figure*}
\begin{minipage}{15cm}

\psfig{file=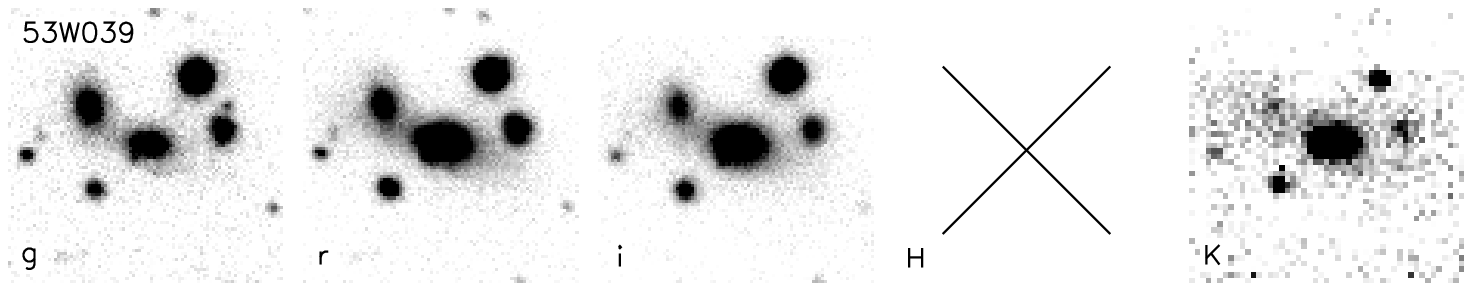}
\psfig{file=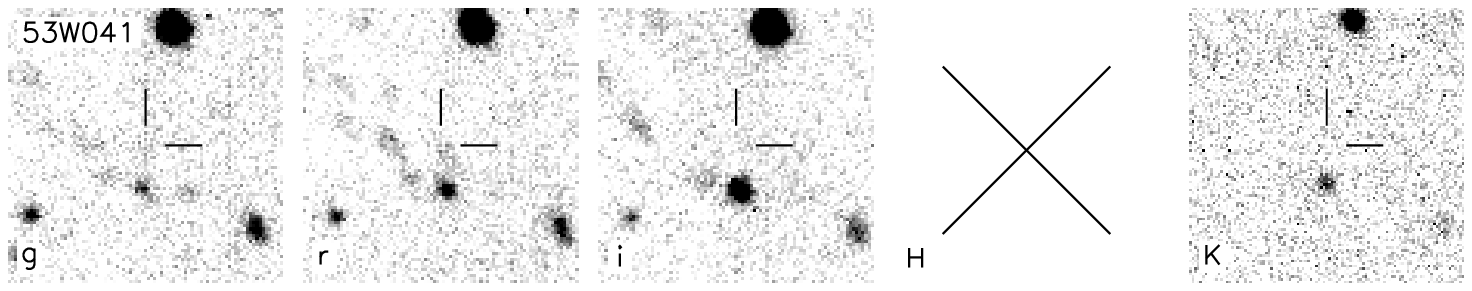}
\psfig{file=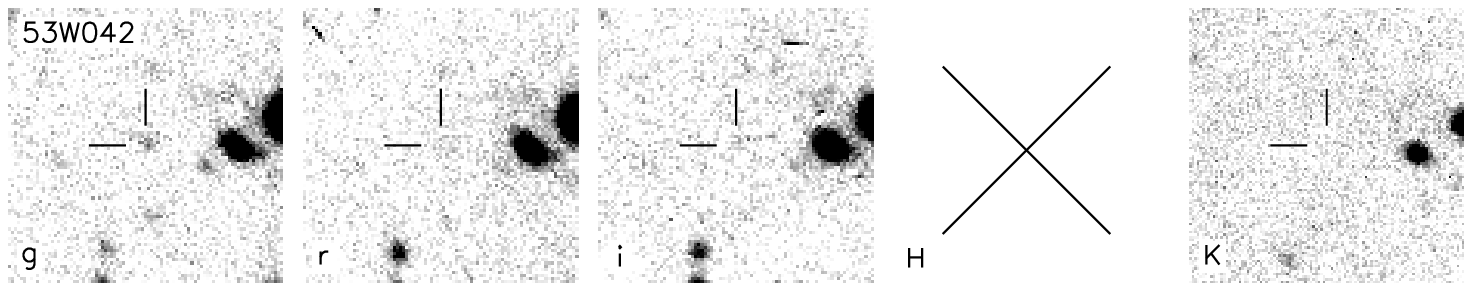}
\psfig{file=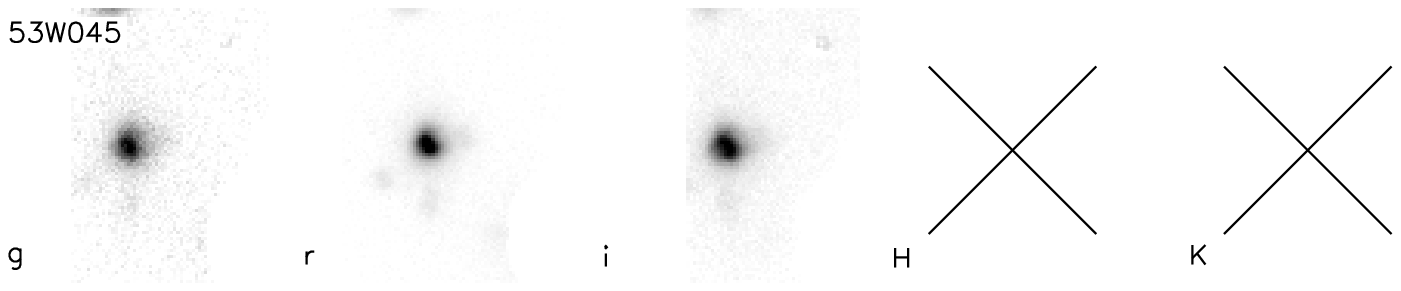}
\psfig{file=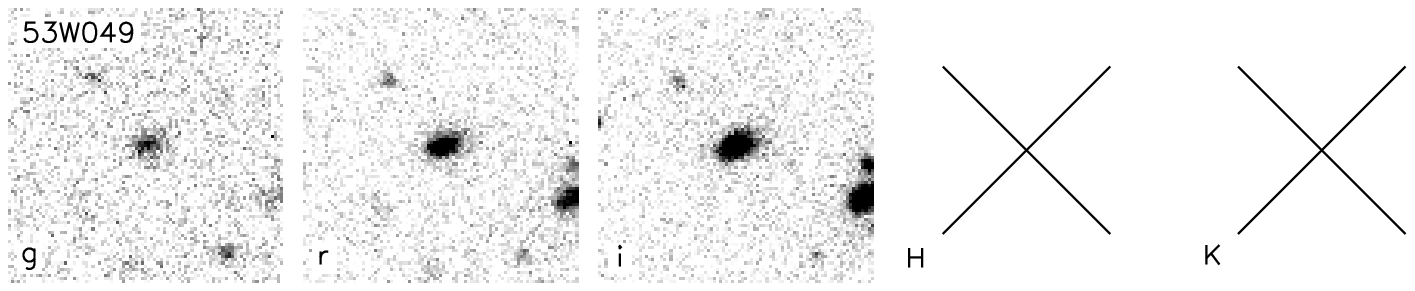}
\psfig{file=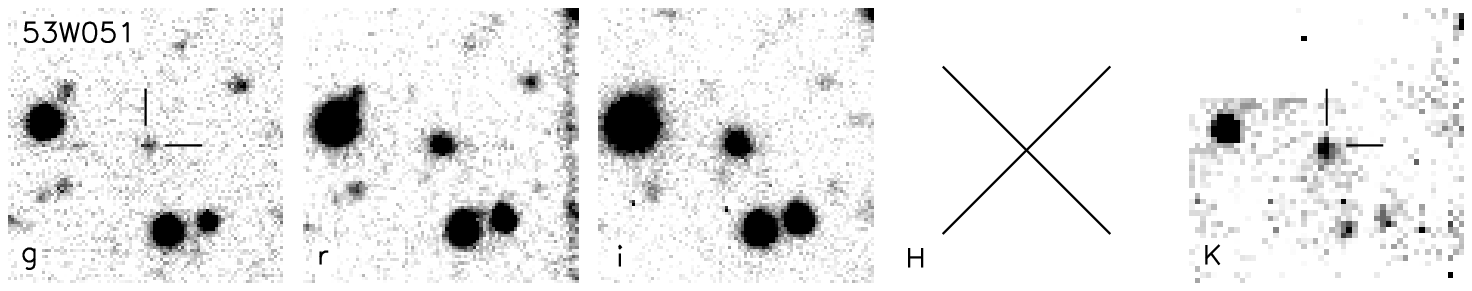}
\psfig{file=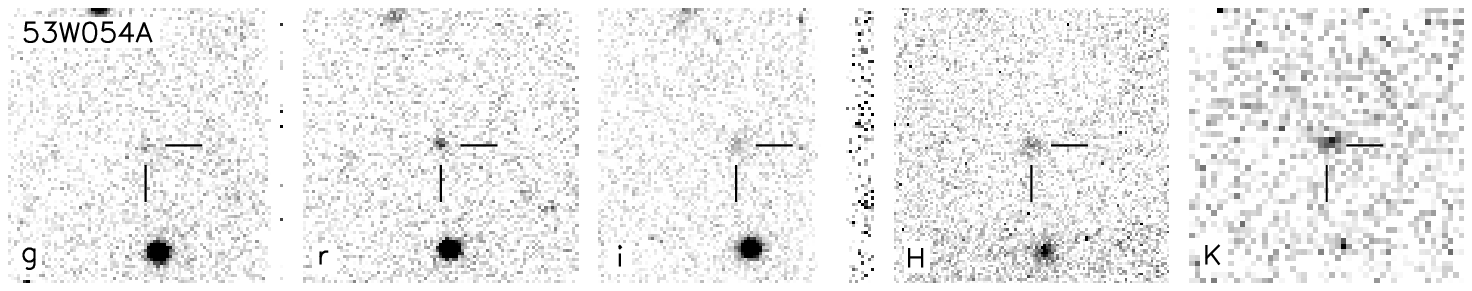}

\contcaption{}
\end{minipage}
\end{figure*}

\begin{figure*}
\begin{minipage}{15cm}

\psfig{file=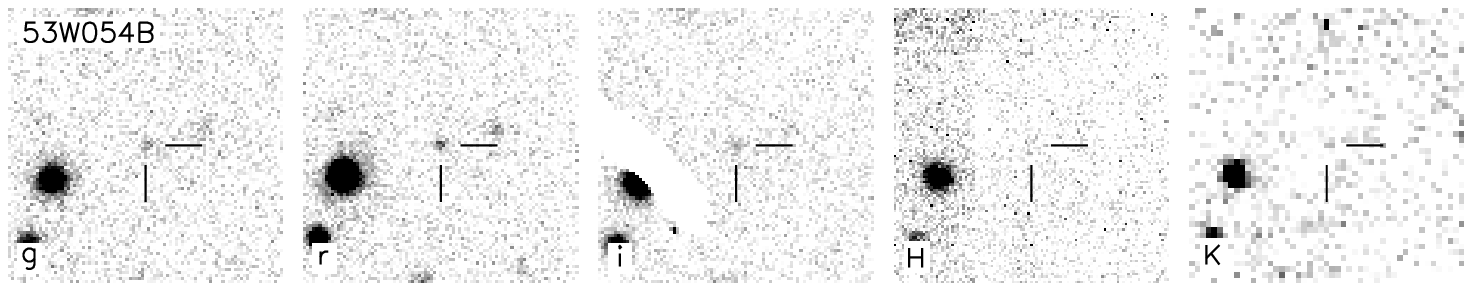}
\psfig{file=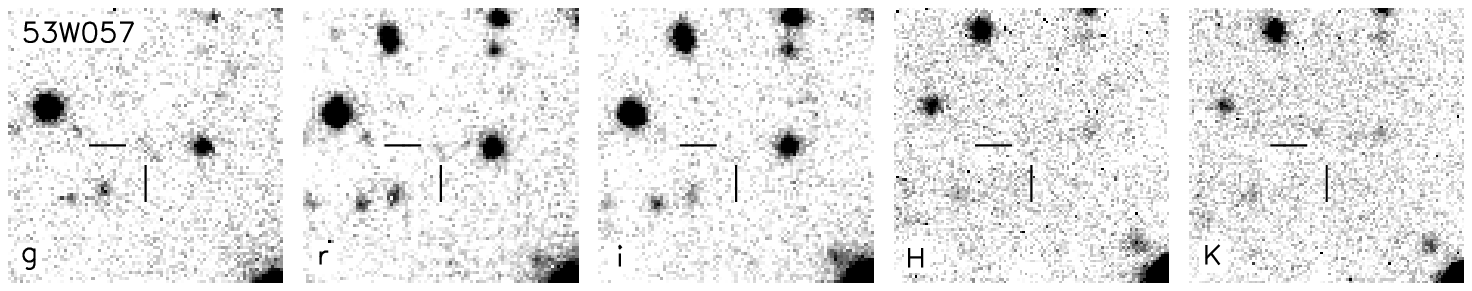}
\psfig{file=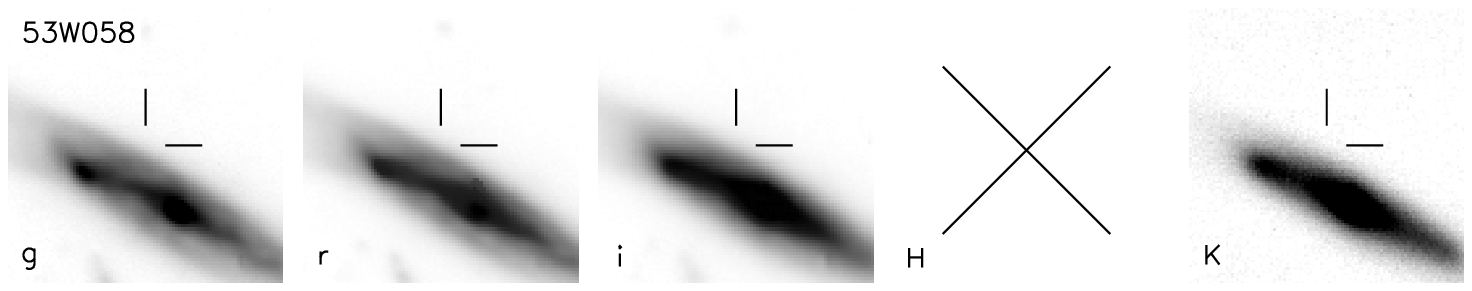}
\psfig{file=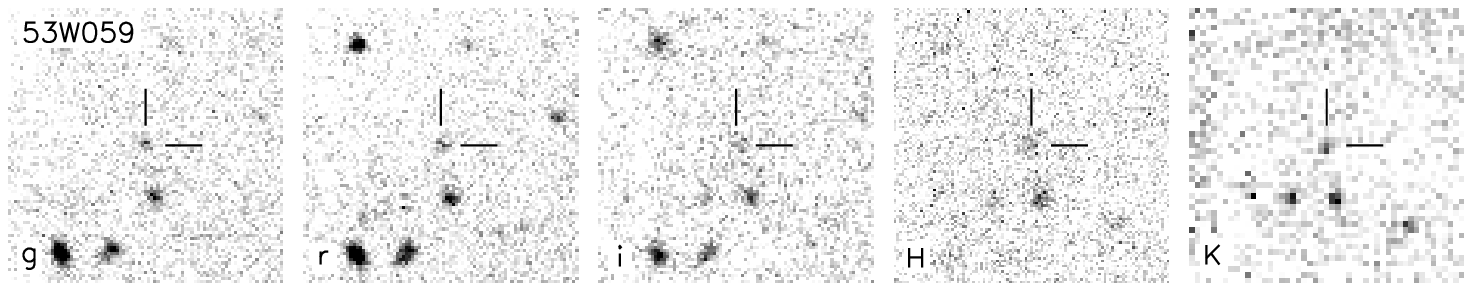}
\psfig{file=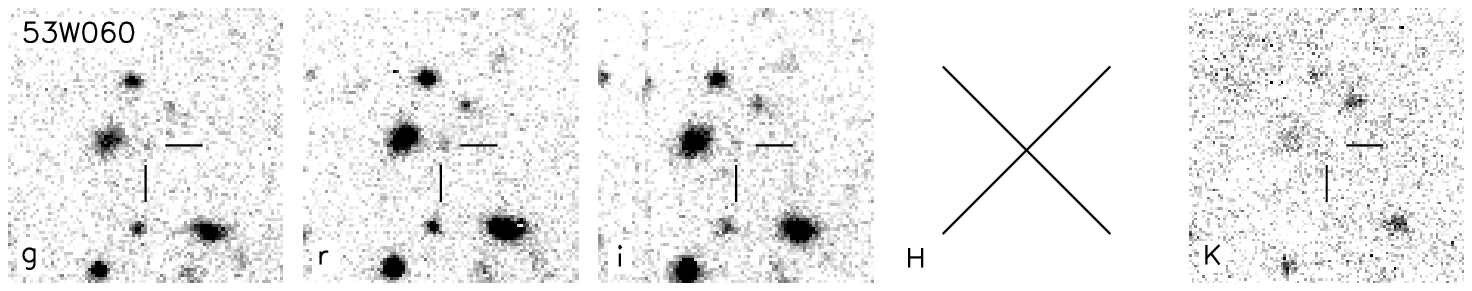}
\psfig{file=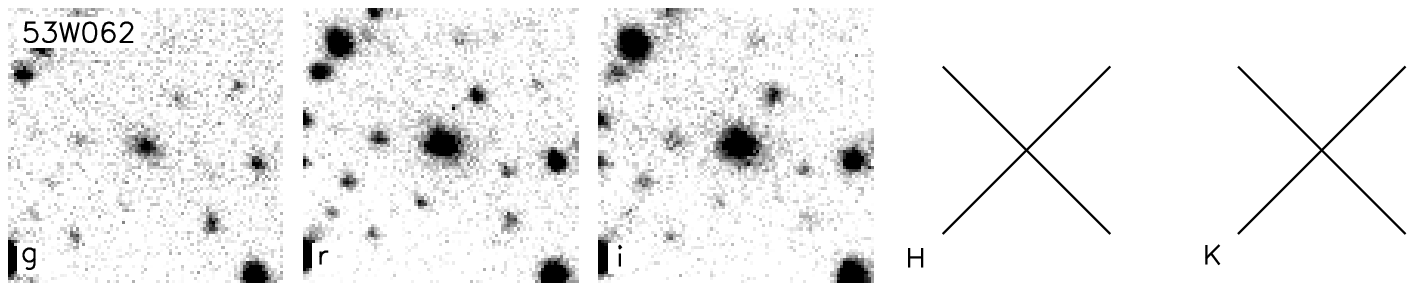}
\psfig{file=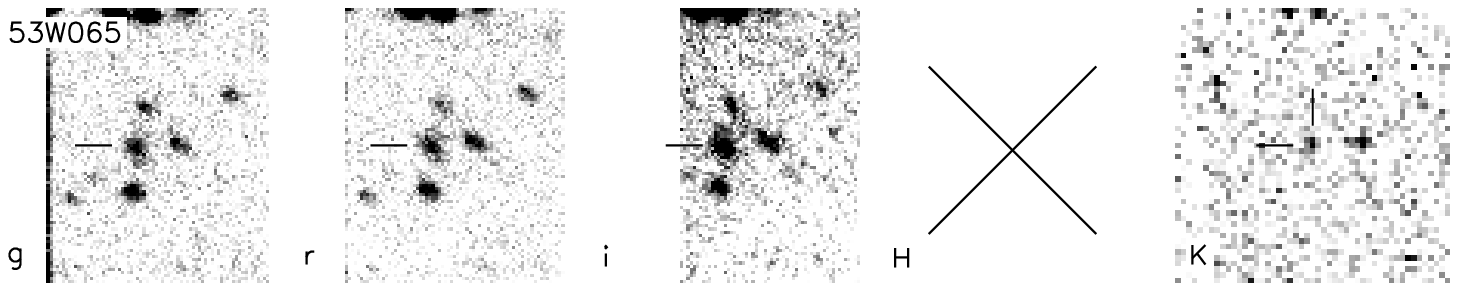}

\contcaption{}
\end{minipage}
\end{figure*}

\begin{figure*}
\begin{minipage}{15cm}

\psfig{file=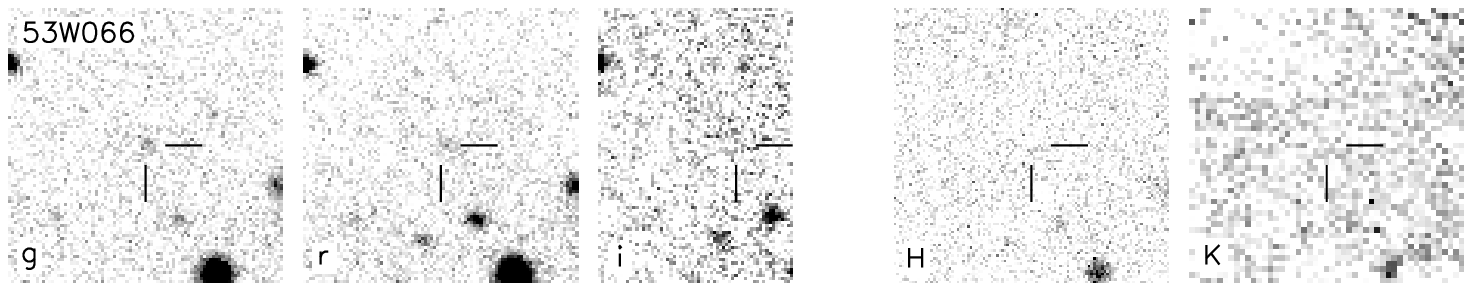}
\psfig{file=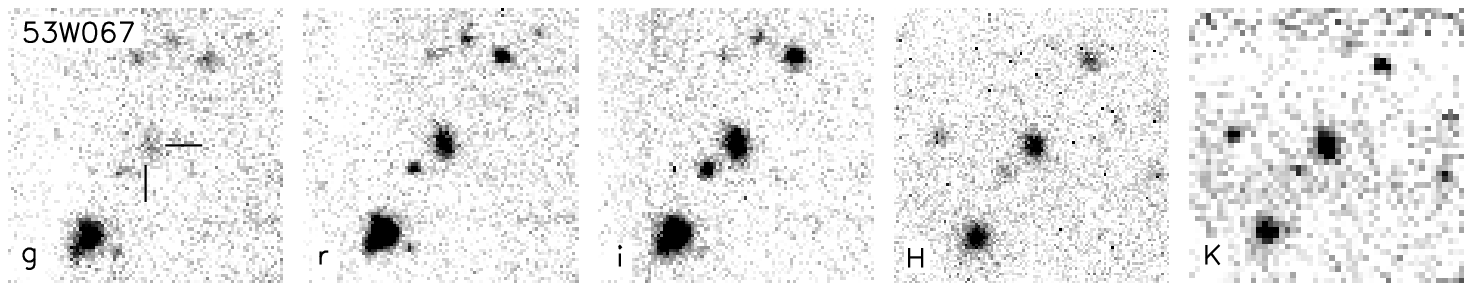}
\psfig{file=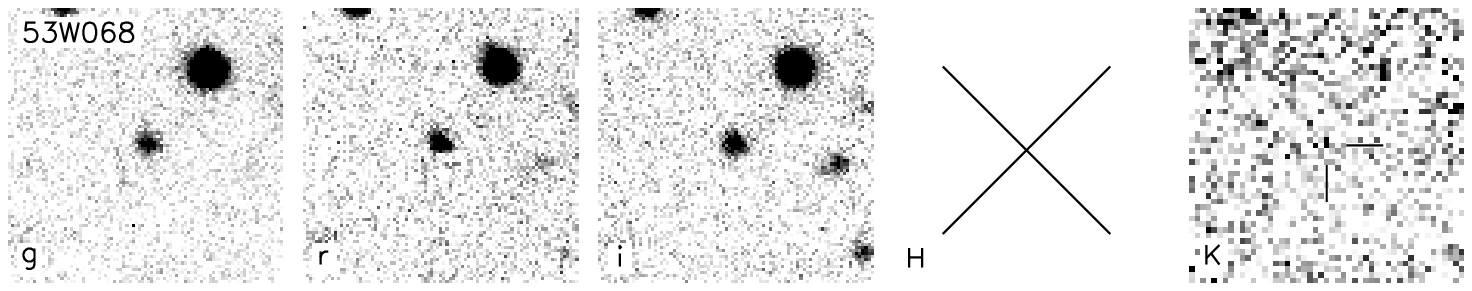}
\psfig{file=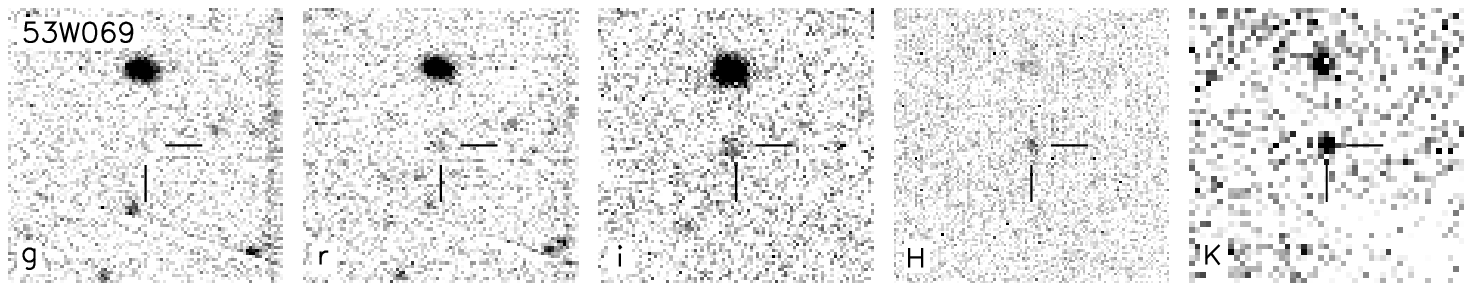}
\psfig{file=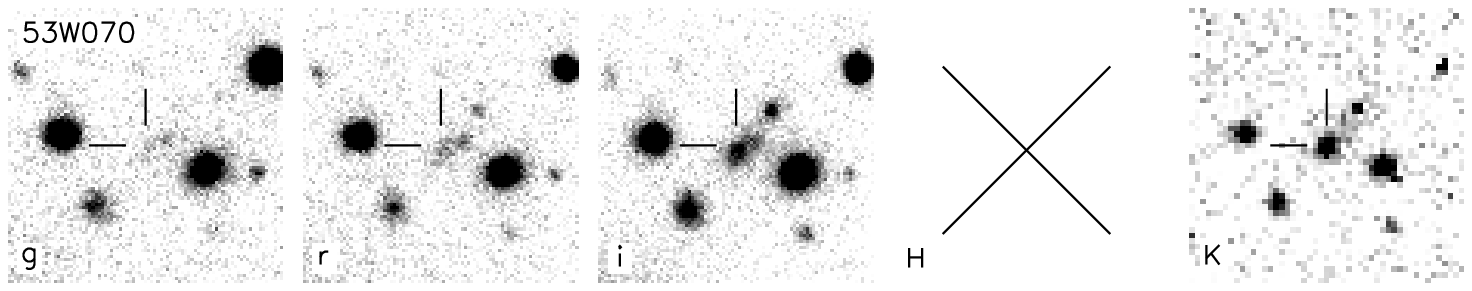}
\psfig{file=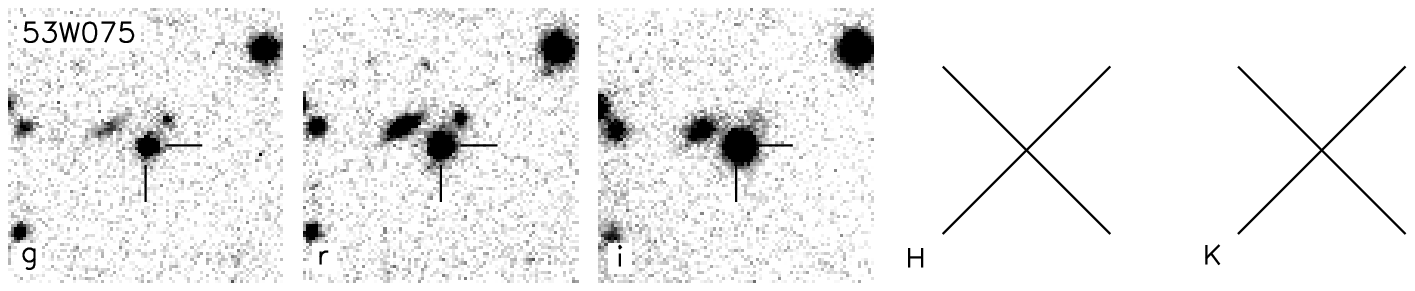}
\psfig{file=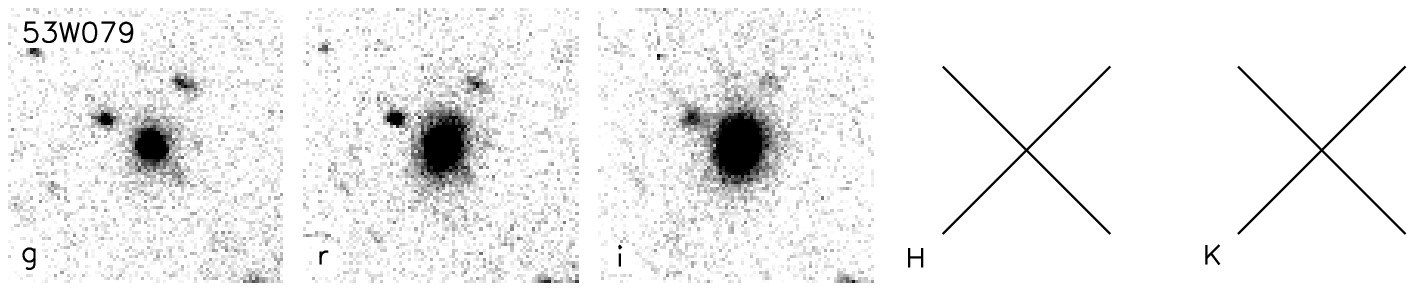}

\contcaption{}
\end{minipage}
\end{figure*}

\begin{figure*}
\begin{minipage}{15cm}

\psfig{file=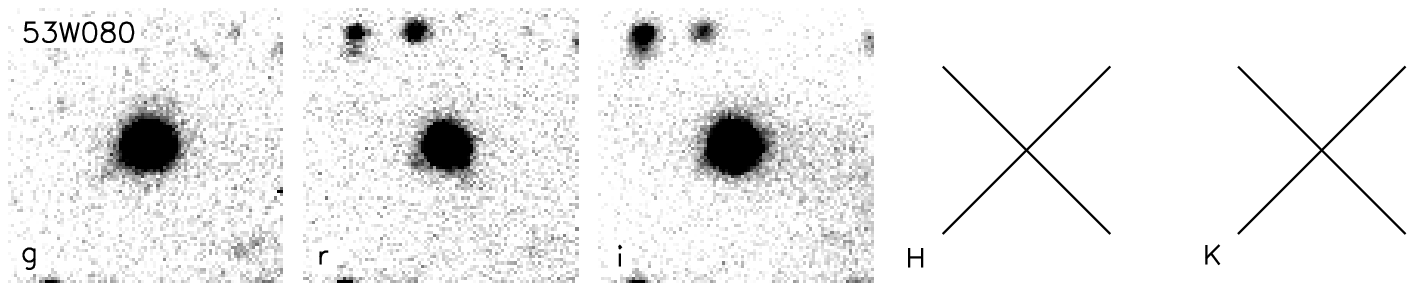}
\psfig{file=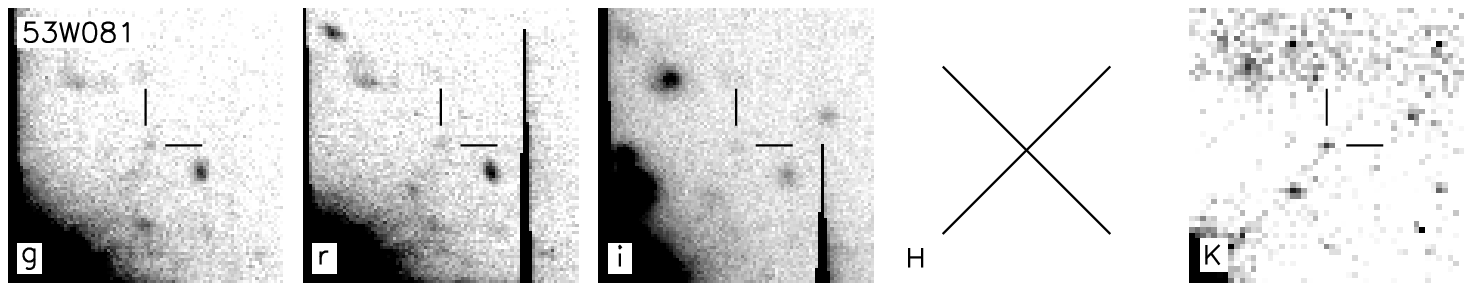}
\psfig{file=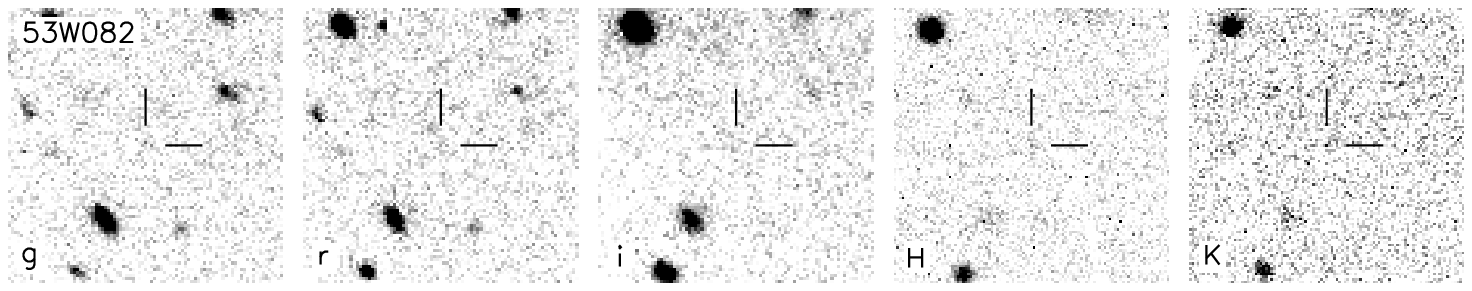}
\psfig{file=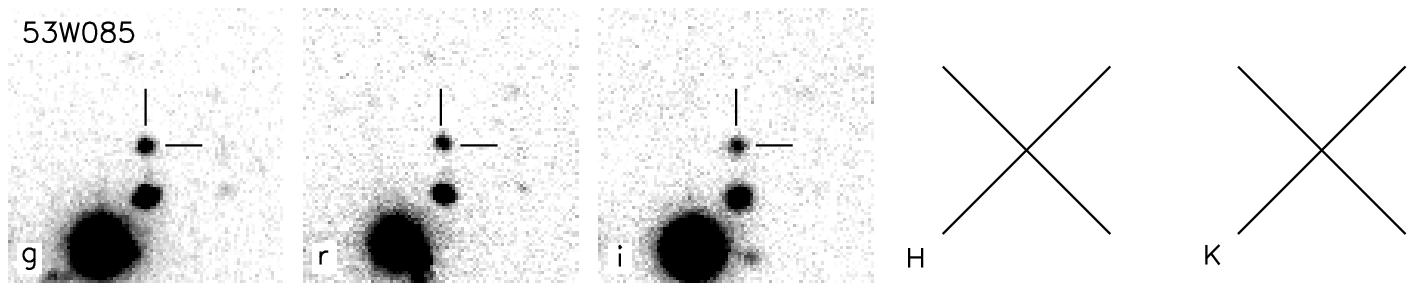}
\psfig{file=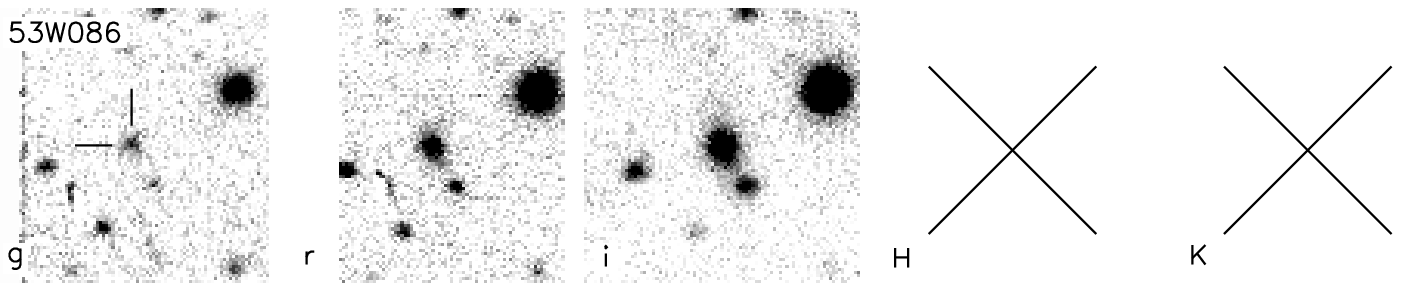}
\psfig{file=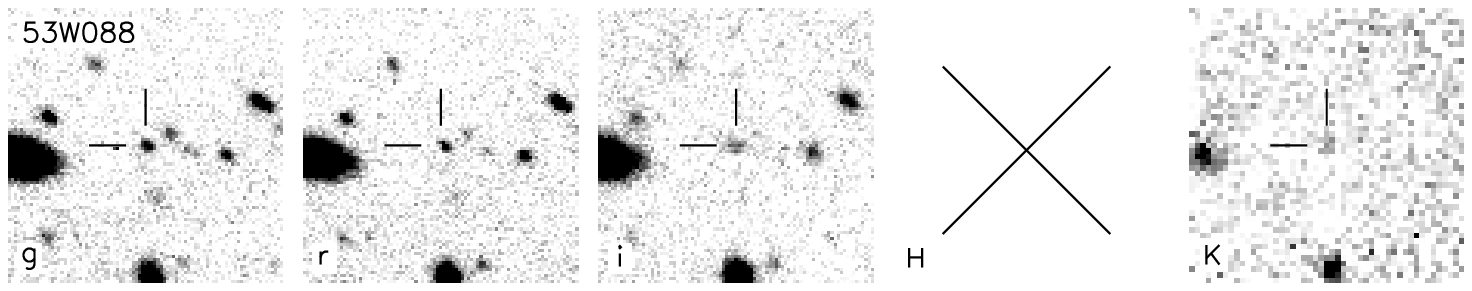}
\psfig{file=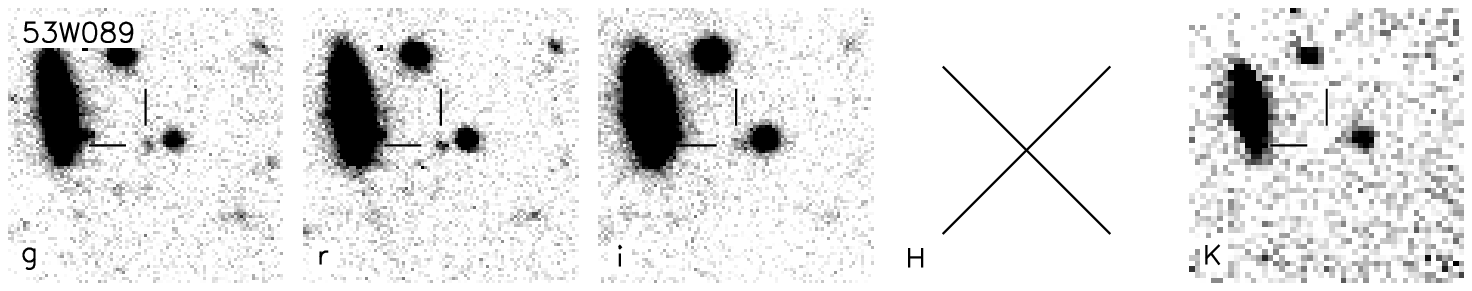}

\contcaption{}
\end{minipage}
\end{figure*}

\begin{figure*}
\begin{minipage}{15cm}

\psfig{file=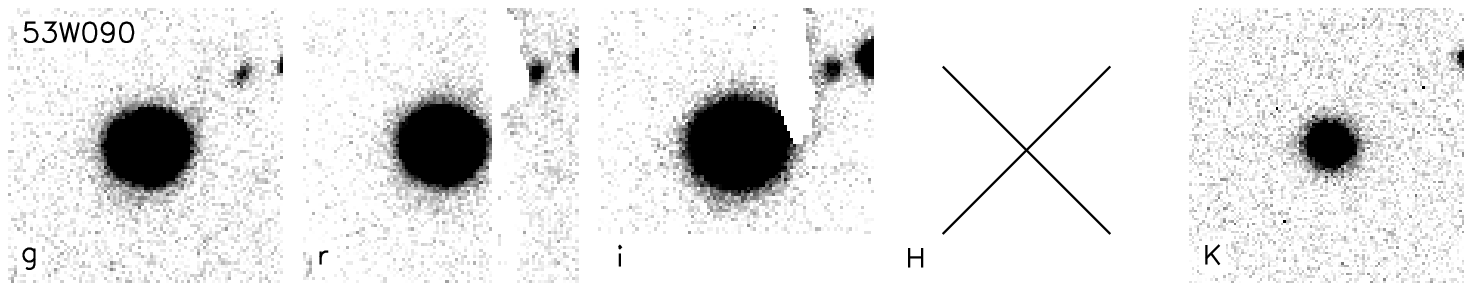}
\psfig{file=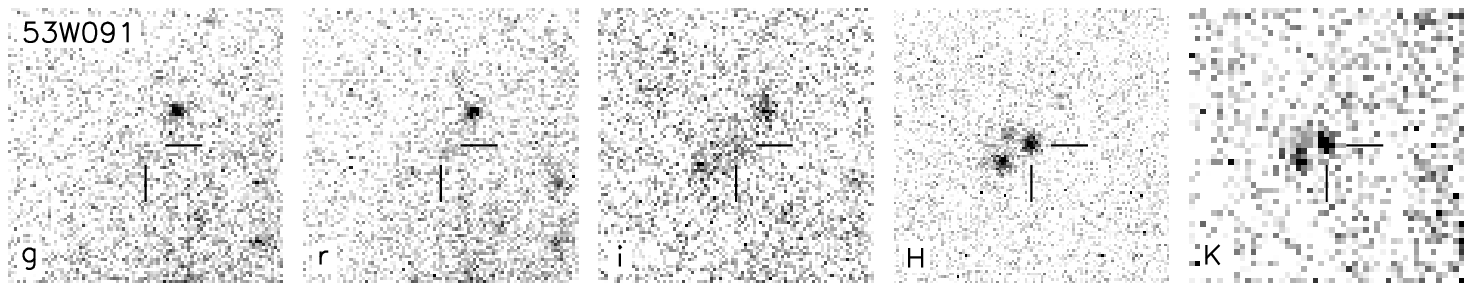}

\contcaption{}
\end{minipage}
\end{figure*}

\subsection{Photometry}

The photometric calibration of the \fsh\ images was done in two stages,
due to the non-photometric conditions of a significant number of
the observing nights.  For most of the sources, at least one exposure
had been taken on a photometric night which could therefore be
calibrated accurately.  This calibrated exposure was then used to
bootstrap a zero-point for the non-photometric mosaic of all the
exposures.

The apparent magnitude of a source ($m$) is defined by 
\begin{equation}
m = m_{\rm inst} + k \cdot \overline{{\rm sec}z} + C \cdot (g-r) +
  z_{\rm p}
\label{mageqn1}
\end{equation}
where $m_{\rm inst}$ is the instrumental magnitude, $k$ is the
extinction coefficient for a given filter, $\overline{{\rm sec}z}$ is
the time-averaged airmass, $C$ is the colour coefficient for the
filter, $(g-r)$ is the instrumental colour of the source, and $z_{\rm
p}$ is the zero-point.  Higher order terms are not included due to the
limited number of standard star observations available.

For each night of the observing runs, there were between three and
eight observations of Gunn standard stars made.  These were taken
throughout the night at different airmasses and in all three filters.
To avoid saturation, the standard stars were observed with the
telescope out of focus.  Instrumental magnitudes in a
30\arcsec--40\arcsec\ circular aperture were measured for each star,
on every night that the log sheets recorded as being photometric.
True apparent magnitudes ($m_0$) were taken from a list of both
published and unpublished standards available at the telescope
\cite{Thuan76,Wade79,Kent85}.  Equation \ref{mageqn1} was then fitted
to the data ($m_0-m_{\rm inst}$, ${\rm sec}z$, $g-r$) using a linear
least squares method.  This gave best-fitting values for $k$, $z_p$
and $C$ for each night.  The colour coefficients are not expected to
change from night to night and the fitted values were indeed
consistent, thus a weighted average over all the nights in a run was
used.  A second least squares fit was then performed with $C$ fixed at
its average value, and values for $k$ and $z_{\rm p}$ were found for
each photometric night.  For the non-photometric mosaics, the
extinction and colour coefficients were taken as the mean of the
values over all the runs (Table~\ref{kccoeffs}).  Note that these
values are consistent with those assumed by Neuschaefer \& Windhorst
(1995)\nocite{Neuschaefer95a}, who used the same instrument and
telescope, over a similar period of time (1984--1988).

For each photometric image containing a radio source, the automated
image detection program {\sc pisa} \cite{Draper96} was used to
calibrate 20--80 sources in the image.  The search parameters were
chosen such that a source was defined as having at least eight
contiguous pixels with values greater than 5-$\sigma$ above the
background.  Such conservative parameters ensured that only good
signal-to-noise sources were detected in the relatively short
photometric exposure.  The corresponding non-photometric, deep mosaic
was then analysed in the same way with {\sc pisa}.  The mean
difference in magnitude between each source, as observed in the mosaic
and in the calibrated exposure, then allowed the zero-point of the
non-photometric mosaic to be bootstrapped from the calibrated
photometric image.


\begin{table}
\caption{Mean extinction \& colour coefficients}
\label{kccoeffs}
\begin{tabular}{ccc}
Filter & $\overline{k}$ & $\overline{C}$\\
 & & \\
$g$  &  $-0.28\pm0.05$  &  $0.16\pm0.02$\\
$r$  &  $-0.14\pm0.03$  &  $0.10\pm0.01$\\
$i$  &  $-0.11\pm0.02$  &  $0.05\pm0.01$\\
\end{tabular}
\end{table}

For five sources (53W010, 53W019, 53W020, 53W022, 53W027), only
$r$-band observations from the 1984 run were available so no colour
correction could be applied to them.  In addition, only two standard
star observations were available for each of the two nights, which was
inadequate to determine both the extinction and zero-points.  Thus,
the mean extinction coefficient from Table~\ref{kccoeffs} was adopted,
and the zero-point calculated from the observed magnitudes of the
standards.  For five other sources (53W080, 53W081, 53W085, 53W086,
53W090), no photometric calibrations were available for the $g$ and
$r$ images.  The zero-points adopted for these mosaics were taken as
the mean of the zero-points for all the other observations in the
sample.  This is reflected in the slightly larger errors for these
sources in Table~\ref{photmaster}.

The sky background subtraction for each source was done using a method
based upon the image reduction package {\sc cassandra} by
D.~P.~Schneider, adapted to run as an {\sc iraf} task using the {\sc
apphot} package for the numerical work.  A rectangular box (typically
10\arcsec--20\arcsec\ on a side) surrounding the source, and excluding
any other objects, was cut out of the mosaic image.  A plane was
fitted to this box, excluding from the fit a circle centered on the
source, and the sky was subtracted from the image.  Photometry was
performed using a circular aperture of 4\farcs0, 7\farcs5 or 10\farcs0
diameter (depending on the extent of the source), with the sky
statistics determined over the remainder of the box.  This enabled a
large asymmetric sky aperture to be used, which is not possible with
the basic {\sc apphot}, in addition to providing a better measure of
the sky background than is possible with an annulus around the object.

During testing of this method an error was found in the way that {\sc
apphot} calculates magnitude errors for faint objects.  {\sc apphot}
makes the approximation $\ln (1+\Delta) \approx \Delta$ which for
faint sources overestimates the error by as much as 0.1--0.2
magnitudes.  This was corrected by including the second, and for good
measure the third, order terms in the expansion: $\ln (1+\Delta)
\approx \Delta - \Delta^2 / 2 + {\Delta^3 / 3}$.  The apparent
magnitude was then calculated from equation~\ref{mageqn1} using the
coefficients in Table~\ref{kccoeffs} and the appropriate zero-points.
The results are presented in Table~\ref{photmaster}.  The errors
quoted are the quadratic sum of the instrumental magnitude errors from
{\sc apphot} (shot-noise in the object aperture, sky-noise), and the
errors in the determination of $k$, $C$ and $z_{\rm p}$.


\begin{figure*}
\psfig{file=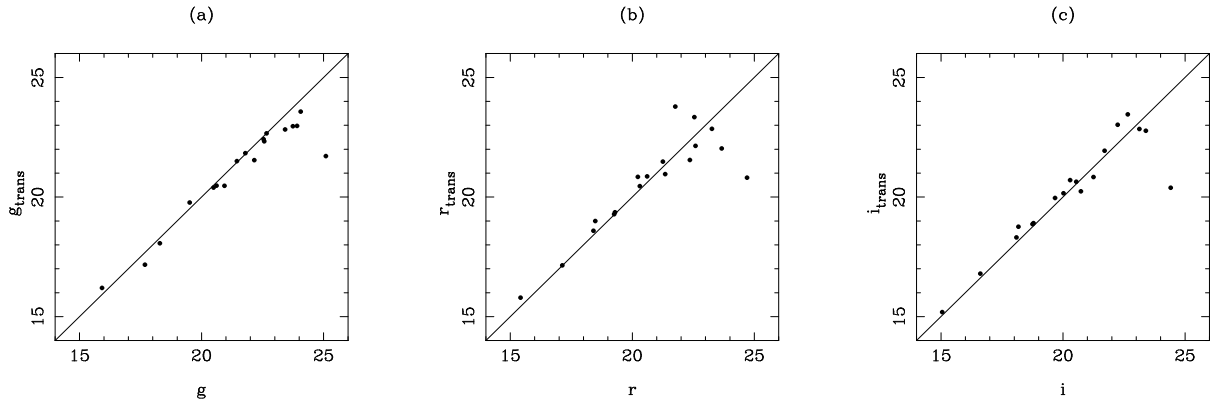,width=160mm}
\caption{Accuracy of the photometric transformations from $J^+F^+N^+$
to $gri$ for the eighteen sources with measurements in both systems.
$g_{\rm trans}$, $r_{\rm trans}$ and $i_{\rm trans}$ are the
photographic magnitudes transformed to the Gunn system with
equations~\ref{gritrans}--\ref{gritrans1}, $g$, $r$ and $i$ are the observed
\fsh\ magnitudes.\label{gritranscomp}}
\end{figure*}

In order to compare those radio sources with photographic magnitudes,
$J^+F^+N^+$, to those sources with \fsh\ Gunn magnitudes, $gri$, it
was necessary to convert one photometric system to another
\cite{Fukugita95}.  Windhorst \etal\ (1991)\nocite{Windhorst91}\
derived transformations from the Gunn system to the photographic
system, and inverting these gives the required equations:
\begin{eqnarray}
\label{gritrans}
g & = & 0.69{J^+} + 0.31{F^+} - 0.19\\
r & = & 1.14{F^+} - 0.14{J^+} + 0.34\\
i & = & {N^+} + 0.75
\label{gritrans1}
\end{eqnarray}
For those sources without \fsh\ observations, the optical magnitudes
in Table~\ref{photmaster} are photographic magnitudes from Kron \etal\
(1985)\nocite{Kron85} transformed to the Gunn system using these
equations.  An error of 0.3 magnitudes is assigned to each of these
values, corresponding to the quadratic sum of the error in the
transformations (0.1~mag) and the error in the photographic magnitudes
(0.2--0.3~mag).  For the eighteen sources with both photographic and
\fsh\ Gunn photometry, the transformed photographic magnitudes are
compared with the \fsh\ data in Fig.~\ref{gritranscomp}.  It can be
seen that the transformations are in good agreement with the data for
$gri \la 22$--23~mag; fainter than this, the $J^+F^+N^+$ magnitudes
are close to the detection limits of the photographic plates and are
less reliable.

\section{Infrared photometry}

Infrared data are crucial to understanding high-redshift galaxies.
Beyond $z \sim 1$, optical observations sample the restframe
ultraviolet emission of galaxies and the restframe optical emission
has shifted into the near-infrared.  Thus in order to properly compare
sources at high-$z$ with those at low-$z$, the analysis must move to
longer wavelengths.  The $K$-band has long been recognized as a
valuable window through which to investigate radio galaxies, and so we
sought to obtain $K$ magnitudes for the complete Hercules sample.

\subsection{Observations with the UK Infrared Telescope}

Approximately half the sample has been observed during three major
observing runs at the 3.8-m UK Infrared Telescope, Mauna Kea, Hawaii
(Table~\ref{hercukirt}).  The first two runs (1992 \& 1993) used the
infrared camera IRCAM, a 62$\times$58-pixel InSb array with a scale of
0.62 arcsec pixel$^{-1}$.  All these observations were done in the
$K$-band.  The 1997 run used the upgraded camera IRCAM3, which had
256$\times$256 pixels with a scale of 0.286 arcsec pixel$^{-1}$.  Deep
$K$ images were obtained for sources which had not previously been
detected, together with $H$-band observations of several galaxies for
which $H$ was considered important for constraining the redshift.
Additional observations of some sources were made during associated
projects and via the UKIRT Service Observing programme.

A standard jittering procedure was used to obtain a median-filtered
sky flat-field simultaneously with the data.  Nine 3-minute exposures
were made, each offset by 8\arcsec\ (1992 \& 1993) or 15\arcsec\ (1997)
from the first position, resulting in a central area around the radio
source with a maximal exposure time of 27 minutes, surrounded by a
border of low signal-to-noise data.  This was repeated two or three
times for the faintest sources.  Observations of faint standard stars
were taken to calibrate the data.

\subsection{Data reduction and calibration}

For all three UKIRT runs, the image processing stage in the reduction was
performed at the telescope by the observers.  This consisted of the
following steps: (i) a dark frame observed prior to the nine-exposure
sequence was subtracted from each image; (ii) a normalised flat-field
was constructed by taking the median of the nine exposures; (iii) each
exposure was then divided by the flat-field; (iv) an extinction
correction was applied; (v) the images were registered and finally
combined into a mosaic after rejection of known bad pixels.

The standard star observations were reduced in the same manner, and
the instrumental magnitude (corrected for extinction) calculated.
Comparison with the known apparent magnitude then yielded a
zero-point.  Standards were observed regularly and the zero-points
were found to vary by $<0.05$~mag throughout each photometric night.
Colour coefficients are small in the infrared and were not determined.

Photometry was performed on each of the mosaics using the same method
as for the optical data (\S 3.5), with the exception of the 1992
observations for which the sky background was determined from the mean
of $\simeq 5$ circular apertures placed randomly around the source.
The results are presented in Table~\ref{photmaster} and
Fig.~\ref{images}.  In addition to the 36 new observations described
here, Table~\ref{photmaster} contains infrared data published by Thuan
\etal\ (1984)\nocite{Thuan84} for sixteen sources, unpublished data
from G. Neugebauer, P. Katgert \etal\ (private communications) for a
further twenty sources, and data on five other galaxies referenced in
the notes to the table.

\section{Discussion of the multi-colour photometry}

The results of the optical and infrared photometry are discussed in
this section, and compared with both earlier results from the LBDS and
with results from other radio surveys.  Discussion of
redshift-dependent properties will be presented in Paper~II, following
analysis of the new spectroscopic data.


\begin{figure}
\psfig{file=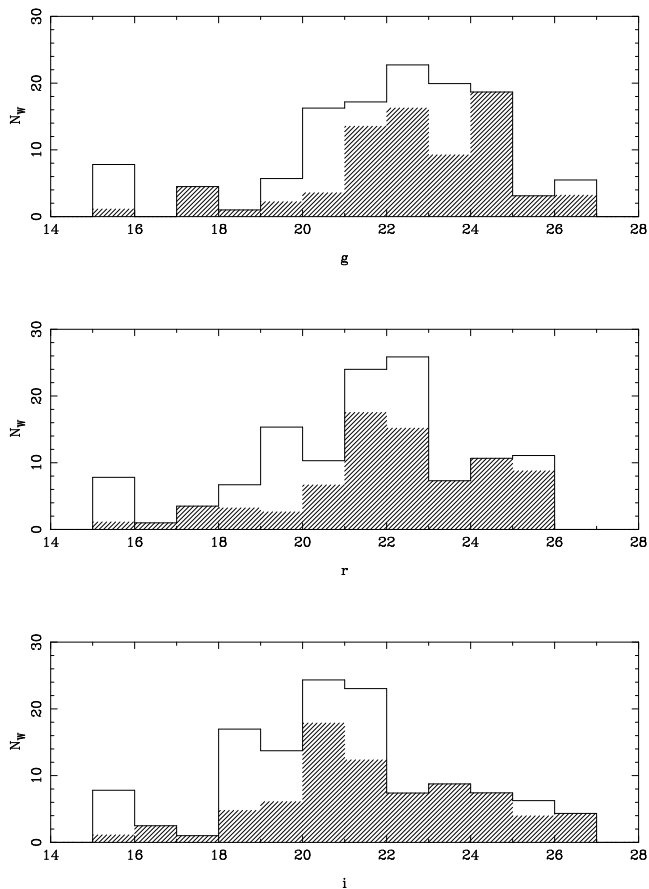,width=85mm}
\caption{Magnitude distributions for the three optical bands ($g$,
$r$, $i$).  Shaded histograms show the 2-mJy sample data, open
histograms show the full Hercules data.  Photographic magnitudes have
been transformed to the Gunn system using equations
\ref{gritrans}--\ref{gritrans1}.  The limits of the photographic data
correspond to $g\simeq 23.0$~mag, $r\simeq 22.3$~mag and $i\simeq
21.5$~mag.\label{grinumbers}}
\end{figure}


\begin{figure}
\psfig{file=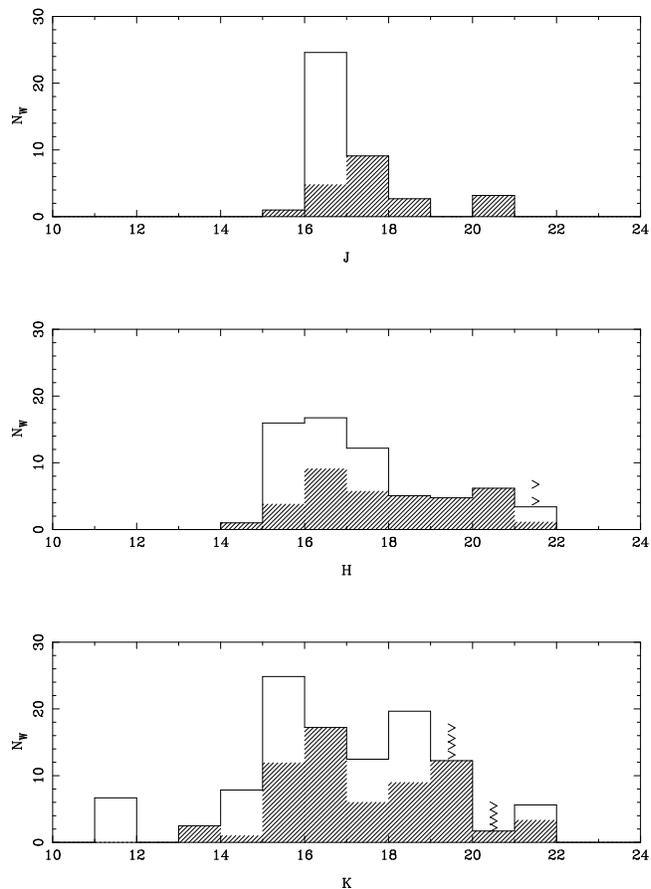,width=85mm}
\caption{Magnitude distributions for the three infrared bands ($J$,
$H$, $K$).  Shaded histograms show the 2-mJy sample data, open
histograms show the full Hercules data.  Arrows denote 3-$\sigma$
upper limits.\label{irnumbers}}
\end{figure}

\subsection{Notes on individual sources}

Before investigating the photometric properties of the sample as a
whole, the optical, infrared \& radio observations of some individual
sources are worth noting.
\begin{description}

\item 53W002 -- This $z=2.390$ radio galaxy has been the subject of
much study (see \S 2.1).  Comparing the optical and infrared colours
recorded in Windhorst \etal\ (1991)\nocite{Windhorst91} with those
here, shows that they are fully consistent within the errors, although
there is some discrepancy between the optical magnitudes.  This may be
explained by noting that Windhorst~et~al.\ use ``total'' magnitudes
which are brighter than the 4\arcsec\ aperture magnitudes in
Table~\ref{photmaster} -- indeed, application of the same method to
these data produces consistent results. It is important to note that
the {\it colours\/} of faint sources are not significantly affected by
aperture size so long as the aperture is larger than the seeing disk,
which it is in all cases here.

\item 53W011 \& 53W013 -- There was originally some concern that these
two sources may have been confused with the grating rings in the WSRT
observations, however both have optical counterparts and so are
retained in the sample.

\item 53W014 -- Optical counterpart is at 3.1-$\sigma$ from the radio
position.  A second object, 5\arcsec\ to the north-east, at
3.7-$\sigma$, may also be a possible identification.

\item 53W022 -- New identification compared with Windhorst \etal\
(1984b)\nocite{Windhorst84b}.  This object looks as if it could be
physically associated with the brighter old identification in the
$r$-band.  At $K$, the two objects are distinct and very similar in
appearance.

\item 53W032 -- A $z=0.370$ galaxy at the centroid of a double radio
source.  The \fsh\ images reveal extended emission to the north-east,
reminiscent of a tidal tail.  Note also that component A is coincident
with an $r\simeq 20.3$~mag source, although component B does not have
an optical counterpart.

\item 53W035 \& 53W089 -- These sources each appeared on two
independent mosaics, enabling the repeatability of the photometry to
be confirmed.

\item 53W054A \& 53W054B -- Originally classified as the lobes of a
classical double radio source \cite{Windhorst84a}, they are now
considered to be two distinct sources.  Both have solid
identifications in the optical, and 53W054A is also identified in the
infrared.

\item 53W057 -- $K$ is a marginal (2-$\sigma$) detection at the
north-east end of the elongated source in $g$.  The brighter source at
6\arcsec\ to the west is also a possible identification, at 2-$\sigma$
from the radio position.

\item 53W058 -- The radio source is in the arm of a bright spiral
galaxy at $z=0.034$.  The coordinates given in Table~\ref{photmaster}
are for the centre of the galaxy; the position of the radio emission
is marked in Fig.~\ref{images}.

\item 53W067 -- The optical counterpart is 4\farcs8 (8-$\sigma$) from
the radio position but has [O{\sc ii}] and Ca H \& K lines identified
in the spectrum, at a redshift of 0.759 (Paper II).  The fainter
source to the south-east, originally proposed as the identification
(at a separation of 1.2-$\sigma$), was identified as an M-type star
from spectroscopy with the Keck telescope (Spinrad, Dey \& Stern,
1997, private communication).

\item 53W069 \& 53W091 -- These two red sources at $z\simeq1.5$ have
been extensively studied with both the Keck Telescope and the {\it
Hubble Space Telescope}.  Their rest-frame ultraviolet spectra are
consistent with an evolved stellar population of ages 3--4~Gyr
\cite{Dey97,Dunlop96,Dunlop99,Spinrad97}, and their rest-frame optical
morphologies are dominated by $r^{1/4}$ elliptical profiles
\cite{Waddington00b}.

\end{description}

\subsection{General properties of the sample}

The apparent magnitude distribution is of cosmological interest
because it is closely related to the radio source redshift
distribution.  Broadly speaking, the fainter a source is (in
optical/infrared wavebands), the more distant it is expected to be.
Although complicated by evolutionary effects, in the absence of
spectroscopic data the magnitude distribution has been an important
tool in studying radio galaxies.

The optical data are shown in Fig.~\ref{grinumbers} and the infrared
data in Fig.~\ref{irnumbers}.  In addition to the full Hercules
distributions, the 2-mJy sample is shown as hatched histograms.  The
first issue that is apparent from these figures is the greater depth
of the new observations.  The limiting magnitudes of the Kron \etal\
(1985)\nocite{Kron85} photographic data correspond to $g\simeq
23.0$~mag, $r\simeq 22.3$~mag and $i\simeq 21.5$~mag.  The \fsh\ data
extend some 3--4 magnitudes fainter in all passbands.  The infrared
$K$-band data are similarly about four magnitudes deeper than the
Thuan \etal\ (1984)\nocite{Thuan84} results.  The $griK$ data are
essential complete, approximately half the sources have $H$-band
measurements and only one-third of the sample has been observed in
$J$.

It is clear from Figs.~\ref{grinumbers} \&~\ref{irnumbers} that the
number of galaxies per magnitude interval does not continue to
increase down to the magnitude limit, but rather turns over at $r \sim
22$~mag.  This immediately points to evolution of some form in the
radio source population -- if the radio sources formed an homogeneous
class of object present throughout the history of the universe, the
numbers would continue to increase towards the magnitude limit.  This
turnover was hinted at in the $N^+$ magnitude distribution of the
whole LBDS \cite{Windhorst84b} and is clearly confirmed for the
Hercules subsample here.  Excluding the 1~mJy~$\le S_{1.4} < 2$~mJy
sources does not significantly change the shape of the magnitude
distributions, but it does {\it increase\/} the median magnitude --
i.e.\ the faintest radio sources are not in general the faintest
sources at optical/infrared wavelengths.  This would suggest that the
faintest radio sources are lower (radio) luminosity objects at lower
redshift, rather than radio-powerful objects at high redshift.  Some
caution must be applied to this conclusion however, given that these
sources are few in number (9) but have large weights ($\ga$2.5).


\begin{figure}
\psfig{file=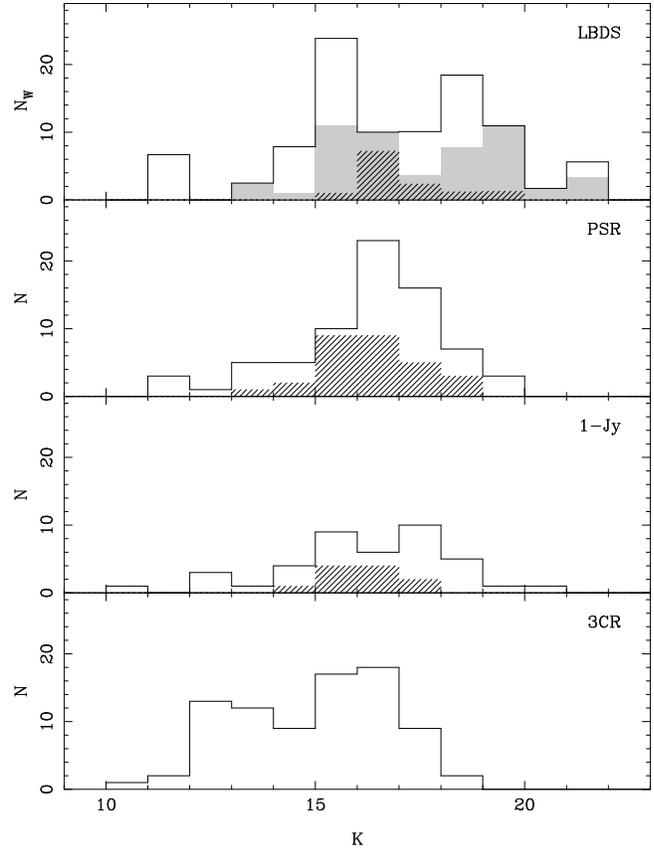,width=85mm}
\caption{Comparison of the $K$-band magnitude distributions for the
LBDS Hercules ($S_\nu \ge 0.001$~Jy at 1.4~GHz), PSR ($S_\nu \ge
0.1$~Jy at 2.7~GHz), 1-Jansky ($S_\nu \ge 1$~Jy at 0.408~GHz) and 3CR
($S_\nu \ge 10$~Jy at 0.178~GHz) samples.  Open histograms are the
galaxies and uncertain IDs (G, G? or ?), hatched histograms are the
quasars (Q or Q?), and the shaded histogram in the top panel is the
2-mJy Hercules sample (note that all the LBDS quasars have $S_{1.4}\ge
2$~mJy).\label{multikband}}
\end{figure}

In Fig.~\ref{multikband}, the LBDS $K$-band magnitude distribution
is compared with the results of three other radio samples with near
complete infrared data.  The Parkes Selected Regions (PSR) data are
principally taken from Dunlop \etal\ (1989)\nocite{Dunlop89a}.  Only
sources in their four-region subsample, for which the $K$-band data
are essentially complete, are used (the 12$^{\rm h}$ and 13$^{\rm h}$
regions are poorly covered in the infrared).  The 1-Jansky sample of
Allington-Smith (1982)\nocite{Allington-Smith82} is a complete sample
of 59 radio sources with 1~Jy~$< S_\nu < 2$~Jy at 408~MHz.  Lilly,
Longair \& Allington-Smith (1985)\nocite{Lilly85} presented infrared
observations of 53 of these sources, which are plotted in the third
panel of Fig.~\ref{multikband}.  Finally, a complete sample of 90
radio galaxies was selected from the 173-source 3CR catalogue of
Laing, Riley \& Longair (1983)\nocite{Laing83}.  Nearly complete
$K$-band data was obtained for this 3CR sub-sample \cite{Lilly84},
which excludes all quasars, and is shown in the last panel of the
figure.  For the first three samples, both galaxies and quasars are
shown.  The LBDS panel also includes the 2-mJy sample for comparison.


\begin{figure*}
\psfig{file=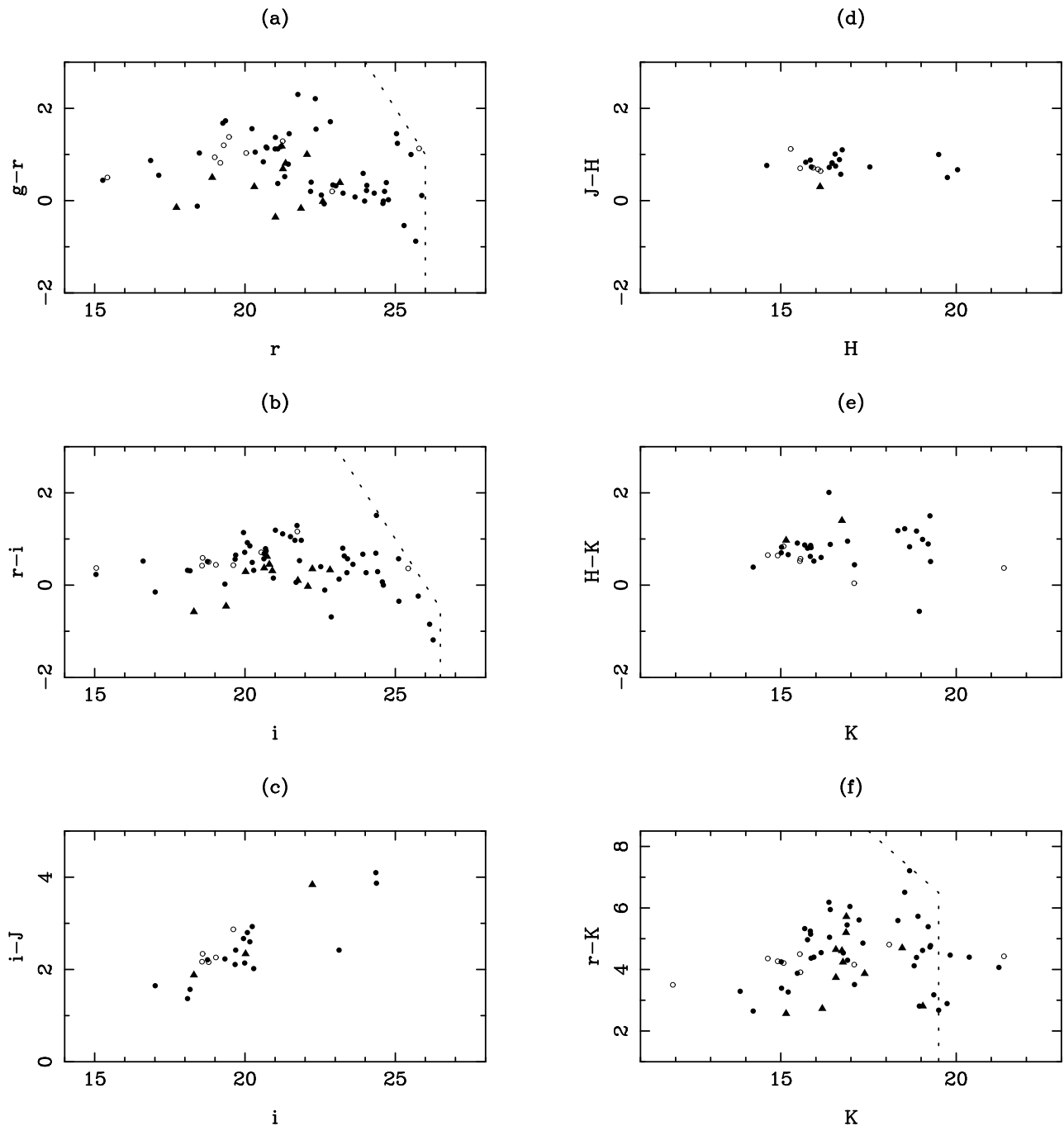,width=170mm}
\caption{Colour--magnitude diagrams for the LBDS Hercules sample.
Solid circles denote galaxies (G, G? or ? types) in the 2-mJy sample.
Open circles denote galaxies with 1~mJy~$\le S_{1.4} < 2$~mJy.  Triangles
denote quasars (Q or Q?), all of which have $S_{1.4} \ge
2$~mJy.  Dashed lines show the completeness limits of the photometry.
The $J$ and $H$ data are not complete to any apparent magnitude, thus
no limits are plotted in (c), (d) and (e).\label{lbdscolmag}}
\end{figure*}

As first pointed out by Dunlop \etal\ (1989)\nocite{Dunlop89a}, the
1-Jansky radio galaxies are biased towards larger $K$ magnitudes than
the (radio) brighter 3CR galaxies.  However, a similar shift is not
seen between the 1-Jansky and PSR sources, a further factor of two
fainter in flux.  Dropping another factor of $\sim$200 in flux, the
LBDS galaxies are seen here to have much the same distribution as the
1-Jansky and PSR galaxies.  The median $K$ magnitudes for the four
samples are 15.3 (3CR), 16.5 (1-Jy), 16.5 (PSR) and 16.8 (LBDS).  To
the extent that the $K$ magnitude of a powerful radio galaxy is a good
estimator of its redshift, this result implies that although the
1-Jansky galaxies are consistent with being high-redshift counterparts
of the 3CR galaxies, these same sources are not seen at even higher
redshifts in the fainter samples.  (Note that a median spectral index
of $\alpha=0.8$ was used to bring the flux limits of the samples to a
common frequency for this comparison.)

Although the number of sources is small, the same effect is not seen
in the quasar distributions.  The median $K$ magnitude of quasars in
the fainter three samples increases with decreasing flux limit: 16.1
(1-Jy), 16.3 (PSR) and 16.8 (LBDS).  This may suggest that the quasar
population is not changing as rapidly as the radio galaxy population.
It must be noted, however, that the decrease in restframe
optical/infrared flux between the 1-Jansky and LBDS quasars (a factor
of $\sim 2$) is certainly not comparable to the drop in radio flux (a
factor of $\sim 400$) -- i.e.\ if the $K$ magnitude is a fair redshift
estimator, the LBDS quasars must be less luminous in the radio than
the 1-Jansky quasars.

Fig.~\ref{lbdscolmag} presents colour--magnitude diagrams for the
Hercules sample.  To compare these results with those of Kron \etal\
(1985)\nocite{Kron85}, note that the colours are related by $(g-r)
\sim (J^+ - F^+) - 0.5$ and $(r-i) \sim (F^+ - N^+) - 0.4$ (from
equations~\ref{gritrans}--\ref{gritrans1}).  It can be seen that the
colours of the new identifications are comparable to those of the
photographic identifications at $r\la 22$~mag.  There is some bias
towards bluer $g-r$ (and to a lesser extent, $r-i$) colours at the
fainter magnitudes.  In particular, there is a deficit of faint
($r>23$~mag), red ($g-r>0.5$, $r-i>1$) radio sources compared with field
galaxy surveys of comparable depth \cite{Neuschaefer92,Neuschaefer95a}.
This is consistent with a continuation of the faint blue radio galaxy
population that Kron \etal\ (1985)\nocite{Kron85} found at brighter
magnitudes.

The sources classified as quasars (Q \& Q?) tend to have bluer optical
colours on average than the galaxies, as one would expect from the
dominance of the AGN at ultraviolet/optical wavelengths in such
sources.  (Recall that the classification was based on morphology, not
colour.)  The ``galaxy'' class includes objects of unknown identity
(?), which dominate at the faintest magnitudes.
Figs.~\ref{lbdscolmag}(a), (b) \& (f) suggest that the bluest of these
objects may in fact be quasars.  Fig.~6 of Dunlop \etal\
(1989)\nocite{Dunlop89a} shows that essentially all PSR sources with
$K \ga 15$ and $R-K \la 3$ are quasars.  Comparing that figure with
Fig.~\ref{lbdscolmag}(f) here, shows that the same cannot be said for
the LBDS -- only one of these faint blue objects is of uncertain
classification (53W068) and one is a probable quasar (53W036), while
the other three are G (53W034, 53W067) and G? (53W027).  The LBDS
quasars have a much broader range of colours than those in the PSR,
perhaps a consequence of their weaker AGN.

Figs.~\ref{lbdscolmag}(c), (d) \& (e) show the distribution of
near-infrared colours with apparent magnitude.  In considering these
figures, it must be remembered how incomplete these data are (only
one-third of the sample has $J$-band data, only one-half has been
observed in the $H$-band).  There appears to be a linear correlation
between $i-J$ colour and $i$ magnitude in Fig.~\ref{lbdscolmag}(c),
however this may simply be due to the absence of faint $J$-band
photometry -- at any given $i$ magnitude it is the reddest sources in
the sample that would have been preferentially detected at $J$.  The
dispersion of the $J-H$ colours (0.2~mag), defined as the standard
deviation about the mean colour, is 2--3 times smaller than the
dispersion in the optical colours.  This is probably due to the
incomplete infrared data, but may also be a genuine property of the
sample.  The smaller scatter may indicate that the infrared emission
is dominated by stellar light from the host galaxies, rather than
emission from the AGN which tends to dominate at bluer wavelengths and
can vary significantly from one source to another.

To summarize, the optical identification of the 72-source LBDS
Hercules sample is essentially complete (only three sources remain
unidentified).  Approximately 85\% of the sources have also been
identified in the near-infrared $K$-band and a programme to obtain
complete $J$- and $H$-band data is ongoing.  Paper~II will present the
results of work to measure redshifts for this sample from both direct
spectroscopy and by using photometric estimation techniques.  This is
now the most comprehensive sample of radio-selected sources available
at millijansky flux density levels, and is ideally suited to study the
evolution of the 1.4~GHz radio luminosity function (for AGN) out to
very high redshifts -- a topic to be addressed in forthcoming papers.

\section*{Acknowledgments}

We thank M. Oort and S. Anderson for contributing to the Palomar
observations; T. Keck for assisting with the processing of the \fsh\
data; and P. Katgert, G. Neugebauer, H. Spinrad and D. Stern for
allowing us to include some unpublished data.  The Hale Telescope at
Palomar Observatory is owned and operated by the California Institute
of Technology.  The United Kingdom Infrared Telescope is operated by
the Joint Astronomy Centre on behalf of the PPARC.  This work was
supported by a PPARC research studentship (to IW); and by NSF grants
AST8821016 \& AST9802963 and NASA grants GO-2405.01-87A,
GO-5308.01-93A \& GO-5985.01-94A from STScI under NASA contract
NAS5-26555 (to RAW).


\bsp

\label{lastpage}

\end{document}